\documentclass[a4paper,11pt]{article}

\pdfoutput=1 

\usepackage{jheppub} 

\usepackage{xspace}

\newcommand{\beq}{\begin{eqnarray}}
\newcommand{\eeq}{\end{eqnarray}}
\newcommand{\pt}{p_{T}}
\newcommand{\ptmin}{p_{T,min}}
\newcommand{\gev}{GeV$^{2}$}
\newcommand{\ba}{\vec{a}_{T}{}}
\newcommand{\bb}{\vec{b}_{T}{}}
\newcommand{\bk}{\vec{k}_{T}{}}
\newcommand{\bq}{\vec{q}_{T}{}}
\newcommand{\bp}{\vec{p}_{T}{}}
\newcommand{\bPhi}{\vec{\Phi}_{T}{}}

\newcommand{\ket}[1]{| {#1} \rangle}
\newcommand{\bra}[1]{\langle {#1} |}
\newcommand{\lpair}{\textsc{lpair}\xspace}
\newcommand{\pptoll}{\textsc{pptoll}\xspace}

\preprint{CP3-14-62}

\keywords{two-photon reaction, $k_t$-factorization, Monte Carlo event generator, 
muon pair production, equivalent photon approximation}

\title{
  Central $\mu^{+}\mu^{-}$ production via photon-photon fusion in proton-proton 
  collisions with proton dissociation
}

\author[a,b]{Gustavo G. da Silveira}
\author[a]{Laurent Forthomme}
\author[a]{Krzysztof Piotrzkowski}
\author[c]{Wolfgang Sch\"afer}
\author[c,d]{and Antoni Szczurek}

\affiliation[a]{Centre for Cosmology, Particle Physics and Phenomenology (CP3), 
                Universit\'e Catholique de Louvain, B-1348 Louvain-la-Neuve, Belgium}
\affiliation[b]{High and Medium Energy Group, Instituto de F\'isica e Matem\'atica,
                Universidade Federal de Pelotas, Caixa Postal 354, CEP 96010-900, Pelotas, RS, Brazil}
\affiliation[c]{Institute of Nuclear Physics PAN, PL-31-342 Krak\'ow, Poland}
\affiliation[d]{University of Rzesz\'ow, PL-35-959 Rzesz\'ow, Poland}

\emailAdd{gustavo.silveira@cern.ch}
\emailAdd{laurent.forthomme@uclouvain.be}
\emailAdd{krzysztof.piotrzkowski@uclouvain.be}
\emailAdd{wolfgang.schafer@ifj.edu.pl}
\emailAdd{antoni.szczurek@ifj.edu.pl}

\abstract{
  We present a formalism which uses fluxes of equivalent photons including 
  transverse momenta of the intermediate photons. The formalism reminds the 
  familiar $k_t$-factorization approach used, e.g., to study the two-photon production of 
  $c\bar{c}$ or $b\bar{b}$ pairs. The results of the new method are compared 
  with those obtained using the code \textsc{lpair}, and a good agreement is 
  obtained. The inclusion of the photon transverse momenta is necessary in studies 
  of correlation observables. We present distributions for the dimuon invariant 
  mass, transverse momentum of the muon pair and relative azimuthal angle 
  between muons separately for elastic-elastic, elastic-inelastic, 
  inelastic-elastic and inelastic-inelastic mechanisms. For typical 
  experimental cuts all mechanisms give similar contributions. The results 
  are shown for different sets of cuts relevant for the LHC experiments. 
  The cross sections in different regions of phase space depend on $F_2$ 
  structure function in different regions of $x$ and $Q^2$. A comment on 
  $F_2$ is made.
}

\begin{document}

\maketitle

\flushbottom

\section{Introduction}
\label{intro}

\begin{figure}[t!]
  \includegraphics[width=1.\columnwidth, bb= 19 40 833 329]{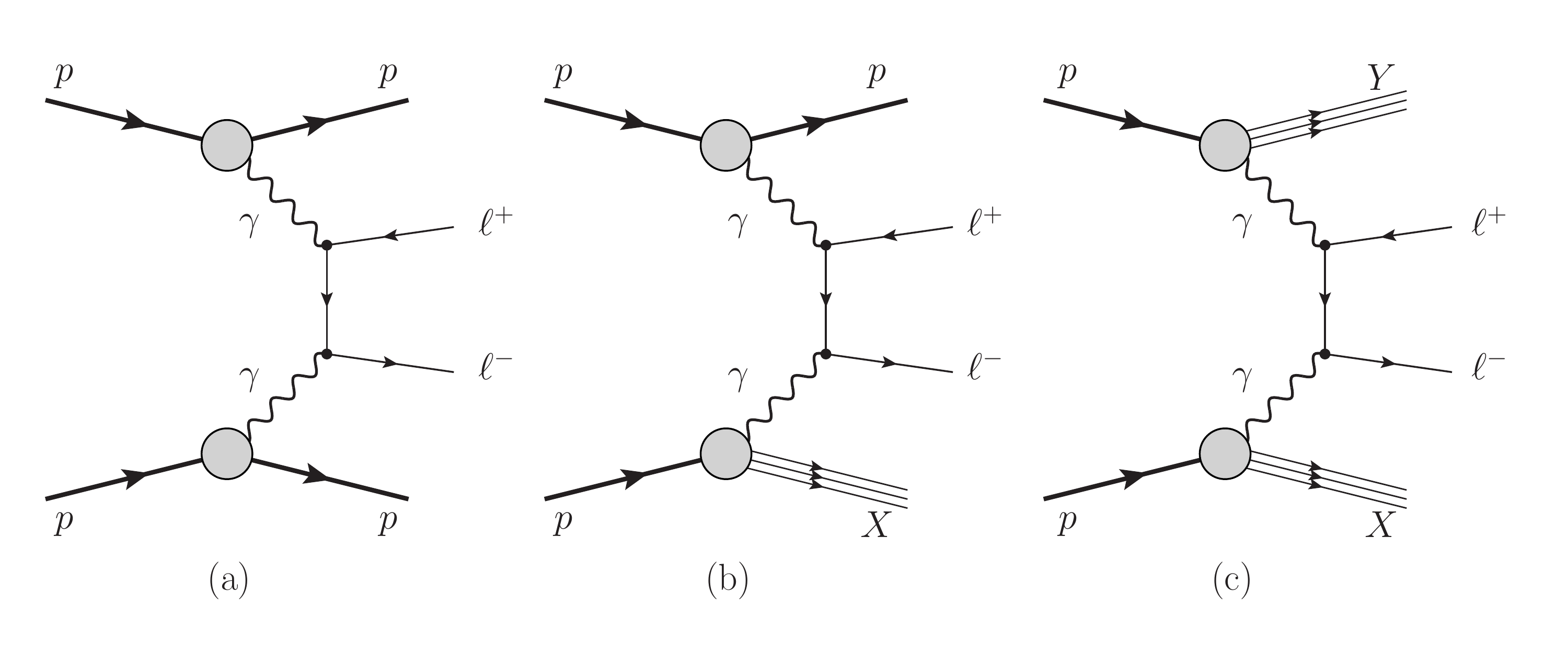}
  \caption{\label{diagrams}
  Diagrams representing the multiperipheral two-photon processes studied in this paper: 
  (a) elastic process, (b) single-dissociative and (c) double-dissociative 
  process. In all three cases it is possible to study lepton pair production, 
  like $e^{+}e^{-}$, $\mu^{+}\mu^{-}$ and $\tau^{+}\tau^{-}$, whereas $X$ 
  and $Y$ represent the hadronic systems resulting from the proton 
  dissociation.
  }
\end{figure}

The production of lepton pairs via $\gamma \gamma$-fusion in hadron-hadron 
collisions has been studied for a long time. Substantial interest in the 
exclusive lepton pair production was generated by the work \cite{Budnev:1973tz},
where it was proposed to use this process, which is 
calculable in QED, to measure the luminosity of a collider. 
However, regarding the investigation proposed in this paper, we are interested 
in the multiperipheral collisions, which are presented in Figure~\ref{diagrams}: 
elastic interaction with both protons intact in the final state 
(Figure~\ref{diagrams}a), one proton dissociates into a hadronic system in the 
final state (Figure~\ref{diagrams}b), and both protons dissociate 
(Figure~\ref{diagrams}c).

Such calculations are typically performed in the equivalent photon 
approximation \cite{budnev}. Here the ingredients are the photon fluxes, which 
are fully specified by the electromagnetic form factors of the colliding 
hadrons, and off-shell cross sections (or density matrices) that describe the 
$\gamma^*(q_1^2) \gamma^*(q_2^2) \to \ell^+ \ell^-$ process. 
It is also straightforward 
to include the breakup of the incoming protons: here the relevant fluxes will 
be calculable in terms of the total (virtual) photoabsorption 
cross sections -- or structure functions -- of the beam particles. 
All the relevant expressions 
for the photon fluxes are given in the review \cite{budnev}, and also certain 
off-shell cross sections can be found there. 

As will be discussed in the present paper, the calculation of inelastic 
unintegrated photon fluxes requires knowledge of the proton structure functions 
in a broad range of $x$ (quark/antiquark longitudinal momentum fraction with 
respect to the proton) and $Q^{2}$ (photon virtuality). In the deep-inelastic 
regime, the structure functions (parton distributions) are related to the proton's 
partonic structure and undergo DGLAP evolution equations. At low virtualities 
the structure function cannot be calculated easily from first principles and 
has to be rather measured. There are some simple models to extend the 
partonic $F_{2}$ to nonperturbative model (e.g., see Ref.~\cite{SU2000}). This model 
nicely describes virtuality dependence of the Gottfried Sum Rule 
\cite{SU2000_GSR}. The very low $Q^{2}$ region was parametrized in 
Ref.~\cite{Fiore:2002re} including pronounced resonance states by fitting data 
from SLAC and JLAB.

In this work we also bring attention to the fact that the relevant formalism 
for $\gamma \gamma$-fusion reactions in the high-energy limit can be understood as 
a type of $k_{T}$-factorization, where the photon fluxes play the role of ''unintegrated'' 
(transverse momentum-dependent) photon densities. Indeed, as will be seen 
below, the cross section takes the exactly analogous form as the 
$k_{T}$-factorization formula for $q\bar{q}$ jet production via gluon-gluon 
fusion (e.g., see Ref.~\cite{Maciula:2013wg}.)

Here we go beyond what is available in the literature by addressing 
distributions in the transverse momentum of the muon pair as well as the 
azimuthal decorrelation of muons. We also use a variety of modern parametrizations 
of the proton structure functions and discuss the uncertainties related to them. 

Another quantitative description of lepton pair production is the \lpair event 
generator \cite{lpair}, which is based on the calculation for two-photon processes~\cite{vermaseren}, 
and also has the possibility to include proton dissociative processes. We 
compare the results of our $k_{T}$-factorization approach to the results obtained 
with \lpair. 

Considering the two-photon production of low- and high-mass systems, this 
work is also motivated by the fact that the experimental results for exclusive 
dimuon production with the CMS detector indicate that the description provided 
by \lpair is not accurate for large transverse momentum of the dimuons \cite{cms-papers,fsq-12-010}. Then, 
in this paper we explore the corresponding phase space using the \lpair and the 
theoretical $k_T$-factorization approach, in order to understand the possible effects that can 
contribute for this discrepancy between data and theory, which can be related 
to the proton structure functions or rescattering corrections not taken into 
account in the event generator.

This paper is organized as follows: in Section~\ref{sec:kt-approach} we describe the 
theoretical approach. Next, the results obtained with the proposed approach 
are presented in Section~\ref{sec:kt-results}. In Sec.~\ref{sec:mc-studies} 
we study the exclusive $\mu^{+}\mu^{-}$ production with the use of \lpair Monte 
Carlo generator. Then, Section~\ref{sec:compare} presents a 
dedicated survey oriented to compare the theoretical predictions. 
Finally, our conclusions are summarized in Section~\ref{concl}.

\section{Lepton pair production including electromagnetic dissociation of 
protons at high energies: a $k_{T}$-factorization approach}
\label{sec:kt-approach}

\subsection{Kinematics}

We want to describe ultrarelativistic collisions involving particles $A,B$ with 
four-momenta $p_A, p_B$ which fulfill the on-shell conditions 
$p_A^2 = m_A^2, \, p_B^2 = m_B^2$. It is useful to introduce the light-like 
momenta:
\begin{eqnarray}
  p_1 = p_A - {m_A^2 \over \tilde s} p_B, \, \, \,   p_2 = p_B - {m_B^2 \over \tilde s} p_A \, ,
\end{eqnarray}
where 
\begin{eqnarray}
  \tilde s = s \cdot { 1 \over 2} \, \left\{ 1 + \sqrt{1 - {4 m_A^2 m_B^2 \over s^2} }\right\} \, , \, s \equiv 2 (p_A \cdot p_B) \, ,
\end{eqnarray}
so that $p_1^2 = p_2^2 = 0$. For an arbitrary four-vector $a$, we then 
introduce the Sudakov-decomposition:
\begin{equation}
  a = \alpha \, p_1 + \beta \, p_2 + a_\perp \, . 
\end{equation}
Correspondingly, we have the Gribov-decomposition of the metric tensor:
\begin{equation}
  g_{\mu \nu} = g_{\mu \nu}^\perp + {p_{1 \mu} p_{2 \nu} +  p_{2\mu} p_{1\nu} \over (p_1 \cdot p_2)} \, .
\end{equation}
The euclidean components of the transverse part of a four vector $a^\perp_\mu \equiv g_{\mu \nu}^\perp \, a_\nu$ will be
denoted by $\vec{a}_{T}$, so that:
\begin{equation}
  (a^\perp \cdot b^\perp ) = - \ba \cdot \bb .
\end{equation}
In the high energy limit adopted by us, we regard as small all quantities of 
the type:
\begin{equation}
  \epsilon \sim  {m^2 \over s}, \, {\bp^2 \over s}, \, {m |\bp| \over s} \, ,
\end{equation}
where $m,\bp$ stand for the  mass and transverse momentum of any of the 
participating particles. In particular, we have that:
\begin{equation}
  s = E^2_{\mathrm {cm}} \, \big( 1 + {\cal{O}(\epsilon)} \big),  \,  2 (p_1 \cdot p_2)  = s \, \big( 1 + {\cal{O}(\epsilon)} \big) \, . 
\end{equation}
Our calculations will be accurate to power accuracy in the small parameter 
$\epsilon$, and below that we will no longer distinguish between quantities 
differing by ${\cal{O}(\epsilon)}$ amounts wherever possible.
 
\subsection{Pair production amplitude}

Our discussion of the pair production amplitude follows, with small notational 
changes, closely the discussion in \cite{Bartos:2001jz}. Let the four-momenta of
leptons be $p_+$ and $p_-$, then, in the Feynman gauge, 
the amplitude takes the form:
\begin{eqnarray}
  M = - i {(4 \pi \alpha_{\mathrm{em}})^2  \over q_1^2 q_2^2} \, V^{A \to X}_\mu(p_A, p_X)   \, 
  g_{\mu \alpha}  \, \bar u_\lambda(p_-) T_{\alpha \beta} v_{\bar \lambda}(p_+) \, \, g_{\beta \nu} \,  V^{B \to Y}_\nu(p_B,p_Y) \, , 
\end{eqnarray}
with
\begin{eqnarray}
  T_{\alpha \beta} = \gamma_\alpha {\hat q_1 - \hat p_+ + m \over (q_1 - p_+)^2 - m^2} \gamma_\beta + 
  \gamma_\beta {\hat q_2 - \hat p_+ + m \over (q_2 - p_+)^2 - m^2} \gamma_\alpha \, ,  \, \hat{q_1} \equiv q_{1\mu} \gamma_\mu \, \mathrm{etc}. 
\end{eqnarray}

The vertex describing the transitions $A \to X$ is:
\begin{eqnarray}
  \sqrt{4 \pi \alpha_{\mathrm{em}}} \, V^{A \to X}_\mu(p_A, p_X) = \bra{X(p_X)} j^{\mathrm{em}}_\mu(0) \ket{A(p_A)} \, ,
\end{eqnarray}
and its square is related to the hadronic tensor:
\begin{eqnarray}
  W_{\mu \nu}^A(q) \equiv \sum_X (2 \pi)^3 \delta^{(4)}(p_X - p_A - q) \,   V^{\dagger A \to X}_\mu(p_A, p_X)  V^{A \to X}_\nu(p_A, p_X) \, .
\end{eqnarray}
Using the Gribov decomposition of the metric tensor, one can show that the 
leading contribution in the high energy limit is obtained by substituting in 
the photon propagators:
\begin{eqnarray}
  g_{\mu \alpha} \to {2 \over s} p_{2\mu} p_{1\alpha} \, , \, g_{\beta \nu} \to {2 \over s} p_{2\nu} p_{1\beta} \, . 
\end{eqnarray}
After this substitution, we obtain for the production amplitude the compact 
``impact factor'' \cite{Frolov:1970ij,Cheng:1970sz} representation:
\begin{eqnarray}
  {\cal{M}}= -i s \, {(8 \pi \alpha_{\mathrm{em}})^2 \over q_1^2 q_2^2} \, N_1(q_1) \, B_{\lambda \bar \lambda}(k_+,k_-;q_1,q_2) \, N_2(q_2) \, ,
\end{eqnarray}
where
\begin{eqnarray}
  &&N_1(q_1) = {1 \over s} p_{2\mu}  V^{A \to X}_\mu(p_A, p_X)
  , \, \,  N_2(q_2) = {1 \over s} p_{1\nu}  V^{B \to Y}_\nu(p_B, p_Y) 
\;  ,  \nonumber \\
  &&B_{\lambda \bar \lambda}(k_+,k_-;q_1,q_2) = {1 \over s} p_{1_\alpha}p_{2\beta} \bar u_\lambda(k_-) \, T_{\alpha \beta} v_{\bar \lambda}(k_+) \, .
\end{eqnarray}
The normalization is the standard one:
\begin{eqnarray}
  d \sigma(AB \to X \ell^+ \ell^- Y) = {1 \over 2 s} \, \overline{|{\cal{M}}|^2} \,  d\Phi(s;\{p_i\}_{i \in X},\{p_j\}_{j \in Y}, k_+,k_-) \, ,  
\end{eqnarray}
The photon $q_1,q_2$ four-vectors have the Sudakov decomposition:
\begin{eqnarray}
  q_i = \alpha_i p_1 + \beta_i p_2 + q_{i\perp} \, ,  
  q_i^2 = \alpha_i \beta_i s - \bq_i^2 \; ,
\end{eqnarray}
where from the on shell conditions of $p_A^2 = m_A^2, p_X^2 = M_X^2, p_B^2 = m_B^2, p_Y^2 = M_Y^2$, we obtain
the small Sudakov parameters
\begin{eqnarray}
  &&\beta_1 =  - { M_X^2 - m_A^2 + \bq_1^2  + \alpha_1 m_A^2 \over (1 - \alpha_1) s} \, , \,  
\alpha_2=  - { M_Y^2 - m_B^2 + \bq_2^2  + \beta_2  m_B^2 \over (1 -
  \beta_2) s} \, , 
\label{Sudakov_parameters} 
\end{eqnarray}
as well as the virtualities of photons
\begin{eqnarray}
  && Q_1^2 \equiv - q_1^2 = {1 \over 1 - \alpha_1} \Big[ \bq_1^2 + \alpha_1 (M_X^2 -m_A^2)  + \alpha^2_1 m_A^2 \Big] \, , \nonumber \\
  && Q_2^2 \equiv - q_2^2 = {1 \over 1 - \beta_2} \Big[ \bq_2^2 + \beta_2 (M_Y^2 - m_B^2) + \beta^2_2  m_B^2 \Big] \, . 
\end{eqnarray}
The cross section differential in the lepton rapidities and transverse momenta can be written in the form
familiar from the $k_T$--factorization:
\begin{eqnarray}
 {d \sigma(AB \to X \ell^+ \ell^- Y) \over dy_+ dy_- d^2\bp_+ d^2\bp_-} &&=  \int  {d^2 \bq_1 \over \pi \bq_1^2} {d^2 \bq_2 \over \pi \bq_2^2}  
 {\cal{F}}^{(i)}_{\gamma^*/A}(x_1,\bq_1) \, {\cal{F}}^{(j)}_{\gamma^*/B}(x_2,\bq_2) 
{d \sigma^*(p_+,p_-;\bq_1,\bq_2) \over dy_+ dy_- d^2\bp_+ d^2\bp_-} \, , \nonumber \\ 
\end{eqnarray}
\label{eq:kt-fact}
where
\begin{eqnarray}
x_1 = {m_{\perp +} \over \sqrt{s}} \, e^{y_+} + { m_{\perp-}  \over \sqrt{s}} e^{y_-} \, , \, 
x_2 = { m_{\perp+} \over \sqrt{s}} \, e^{-y_+} + { m_{\perp-} \over \sqrt{s}} e^{-y_-} \, , \, m_{\perp \pm} = \sqrt{\bp_\pm^2 + m_\ell^2} \, .
\nonumber 
\end{eqnarray}
The unintegrated photon densities $ {\cal{F}}^{(i)}_{\gamma^*/A}(x,\bq_2) $, where $(i = \mathrm{el,inel})$ read explicitly:
\begin{eqnarray}
  {\cal{F}}^{(\mathrm{el})}_{\gamma/A}(\alpha,\bq^2) = {\alpha_{\mathrm{em}} \over \pi} (1 - \alpha) \, \left[ {\bq^2 \over \bq^2 + \alpha^2 m_A^2} \right]^2 \, 
  {4m_p^2 G_E^2(Q^{2}) + Q^{2} G_M^2(Q^{2}) \over 4m_p^2 + Q^{2} } \,  \left( 1 - {Q^{2} - \bq^2  \over Q^{2}} \right) \, , \nonumber \\
\end{eqnarray}
\begin{eqnarray}
  {\cal{F}}^{(\mathrm{inel})}_{\gamma/A}(\alpha,\bq^2) = {\alpha_{\mathrm{em}} \over \pi} (1 - \alpha) \int^\infty_{M^2_{\mathrm{thr}}} 
  {dM_X^2F_2(M_X^2,Q^{2})  \over M_X^2 + Q^{2} - m_p^2}  \Big( 1 - {Q^{2} - \bq^2  \over Q^{2}} \Big) 
  \nonumber \\
  \times \left[ {\bq^2 \over \bq^2 + \alpha (M_X^2 - m_A^2) + \alpha^2 m_A^2} \right]^2 
  \, .
  \nonumber \\
\end{eqnarray}
Notice that it is a property of the high-energy limit that fluxes of transverse and longitudinal virtual photons are equal and
only the structure function $F_2 = 2x F_1 + F_L$ appears. 

\subsection{Off-shell cross section}

\subsubsection{$\gamma^* \gamma^* \to \ell^+ \ell^-$}
The second ingredient of our calculation is the off-shell cross section for the fusion of virtual photons,
which reads\footnote{Note that (the transverse) virtual photons carry the linear polarizations parallel to $\bq_1, \bq_2$, and the
averaging over photon polarizations is in fact effected by the azimuthal integrations in Eq.(\ref{eq:kt-fact}).}:
\begin{eqnarray}
{d \sigma^*(p_+,p_-;\bq_1,\bq_2) \over dy_1 dy_2 d^2\bp_+ d^2\bp_-}   = 
{\alpha^2_{\mathrm{em}} \over \bq_1^2 \bq_2^2 } \, \sum_{\lambda, \bar \lambda} \, \Big| B_{\lambda \bar\lambda } (p_+,p_-;q_1,q_2) \Big|^2 \, \delta^{(2)}(\bq_1 + \bq_2 - \bp_+ - \bp_-). 
\nonumber 
\end{eqnarray}

We parametrize the lepton four-momenta in terms of their Sudakov-parameters as:
\begin{eqnarray}
  p_\pm = \alpha_\pm p_1 + \beta_\pm p_2 + p_{\pm\perp},  \, \beta_\pm = {\bp^2_\pm + m^2 \over \alpha_\pm s} \, .
\end{eqnarray}
Then, for the off-shell cross section, a particularly simple form can be obtained in terms of the variables \cite{Bartos:2001jz}:
\begin{eqnarray}
z_\pm = {\alpha_\pm \over \alpha}\, , \, \bk = z_- \bp_+ - z_+\bp_- \, . 
\end{eqnarray}
The familiar structures
\begin{eqnarray}
&&\Phi_0 = { 1 \over (\bk + z_+ \bq_2)^2 + \varepsilon^2 } -  { 1 \over (\bk - z_- \bq_2)^2 + \varepsilon^2 } \, , \nonumber \\
&&\bPhi = { \bk + z_+ \bq_2  \over (\bk + z_+ \bq_2)^2 + \varepsilon^2 } -  { \bk - z_- \bq_2 \over (\bk - z_- \bq_2)^2 + \varepsilon^2 } \, ,
\end{eqnarray}
with $\varepsilon^2 = m_\ell^2 + z_+ z_- + Q_1^2$,
enter the off-shell matrix element (see e.g. \cite{Bartos:2001jz} \footnote{We correct a typo in the longitudinal photon contribution in \cite{Bartos:2001jz}.}.):
\begin{eqnarray}
  \sum_{\lambda, \bar \lambda} \, \Big| B_{\lambda \bar\lambda } (p_+,p_-;q_1,q_2) \Big|^2
&&= { 1 \over s^2}  \sum_{\lambda, \bar \lambda} \Big| p_{1_\alpha}p_{2\beta} \bar u_\lambda(p_-) \, T_{\alpha \beta} v_{\bar \lambda}(p_+) \Big|^2 
\nonumber \\
&&= 2 z_+ z_- \bq_1^2
\Big[ {\underbrace{ 4 z_+^2 z_-^2 \bq_1^2  \Phi_0^2}_{L}}  \nonumber \\
&& \hspace{4.6em}+~ {\underbrace{(z_+^2 + z_-^2) \bPhi^2 + m_\ell^2 \Phi_0^2}_{T}} \nonumber \\
&& \hspace{4.6em}+~ {\underbrace{ \Big[ \bPhi \times {\bq_1 \over |\bq|} \Big]^2 - \Big( {\bPhi \cdot \bq_1 \over |\bq_1|}\Big)^2}_{TT'}}
 \nonumber \\
&& \hspace{4.6em}+~ {\underbrace{ 4 z_+ z_- (z_+ - z_-) \Phi_0 (\bq_1 \bPhi)}_{LT}} \Big] \, .
\end{eqnarray}
Here we indicated the terms corresponding to photon $1$ being in the longitudinal or transverse polarization
states and the respective interference contributions. 

Some comments on the applicability of the proposed formalism are in order.

Firstly, limitations due to energy-momentum conservation can be important
in some regions of the phase space.
We impose explicitly the kinematical limitations on the masses of the produced 
dissociative systems
\begin{eqnarray}
W_{\gamma_1 p_2} &=& \sqrt{s_{\gamma_1 p_2}} = \sqrt{x_1 s - \bq_1^2}
> M_Y + M_{\ell \ell} > M_Y  \; , \nonumber \\
W_{\gamma_2 p_1} &=& \sqrt{s_{\gamma_2 p_1}} = \sqrt{x_2 s - \bq_2^2}
> M_X + M_{\ell \ell} > M_X  \;  
\label{limits_on_MX_or_MY}
\end{eqnarray}
for elastic-inelastic (first), inelastic-elastic (second) or
inelastic-inelastic (both) contributions. Here the invariant mass of the dilepton system is
given by
\begin{eqnarray}
M^2_{\ell \ell} =  m_{\perp +}^2 + m_{\perp -}^2 + 2 m_{\perp +}m_{\perp -}\,\cosh(y_+ - y_-) - (\bp_+ + \bp_-)^2\, .
\end{eqnarray}
In practice the phase-space constraint of Eq.~\ref{limits_on_MX_or_MY} is very weak, as
the distributions of interest are falling sufficiently fast.

Secondly, the evaluation of off-shell matrix elements explicitly uses the smallness of
Sudakov parameters given in Eq.~\ref{Sudakov_parameters} in the expansion of the photon momenta.
In order to be able to approximate e.g. $q_1 = x_1 p_1 + q_{1\perp}$, one should make sure
that the condition
\begin{eqnarray}
x_1 \gg { M_X^2 - m_A^2 + \bq_1^2  + x_1 m_A^2 \over (1 - x_1) s}
\label{Sudakov_constraint}
\end{eqnarray}
is fulfilled (a similar constraint with $1 \leftrightarrow 2, M_X \leftrightarrow M_Y$ must hold for $x_2$).
This constraint in practice means that at a given center-of-mass energy, one should not go to too 
large invariant masses of the proton remnant. 
For photons coupling to an elastic transition the form factor cutoff
on $\bq_1^2$ is strong enough to ensure that Eq.~\ref{Sudakov_constraint} is very well fulfilled
in the energy regime of interest.
 
\subsection{Decorrelation of leptons}

The $k_T$-factorization formulation is especially well suited to the description
of pair production in the region of phase space where the total transverse
momentum $\bp_{\textrm{sum}} = \bp_+ + \bp_-$ of the lepton pair is non vanishing.
In the case of intact protons in the final state, the back-to-back correlation
of leptons is very strong and in fact an equivalent photon approximation with
collinear photons would have been of sufficient accuracy.

The situation is different however in the processes with inelastic excitation 
of the proton, where a large transverse momentum can be transferred via the
photon exchange.

Let us look at the situation when one of the protons stays intact, while
the other one dissociates. We can then obtain a very simple form for the
cross section in the limit that the decorrelation momentum $\bp_{\textrm{sum}}$ 
is much larger than the cutoff in the elastic form factor 
$\Lambda \sim 0.7 \, \, \mathrm{GeV}$.
In this limit, we can neglect inside the matrix element $B$ the transverse momentum
of the photon coupling to the ``elastic'' proton and hence integrate out one of the 
photon transverse momenta. The decorrelation momentum $\bp_{\textrm{sum}}$ is then exactly equal
to the  transverse momentum carried by the second photon.
Neglecting lepton masses (which is anyway appropriate in the kinematic region of 
interest in this paper), we obtain the compact form for the cross section
\begin{eqnarray}
  {d \sigma (AB \to A \ell^+ \ell^- X) \over dy_+ dy_- d^2\bp_+ d^2\bp_-} = n(x_1) \, 
{\alpha_{\mathrm{em}}^2 {\cal{F}}^{(\mathrm{inel})}(x_2, \bp_+ + \bp_-) \over (\bp_+ + \bp_-)^2} \, {2 z_+ z_- (z_+^2 + z_-^2)\over \bp_+^2 \bp_-^2} \, .
\end{eqnarray}
Here
\begin{eqnarray}
 n(x_1) = \int {d^2 \bq_1 \over \pi \bq_1^2} \, {\cal{F}}^{(\mathrm{el})}(x_1,\bq_1) \,
\end{eqnarray}
is the Weizs\"acker-Williams flux of photons in the elastically scattered proton.
This result is fully analogous to the decorrelation momentum distribution of $q\bar q$ dijets 
in the photon-gluon fusion obtained in \cite{Szczurek:2000pj}.
\section{Results of $k_{T}$-factorization approach}
\label{sec:kt-results}

In this section we present results obtained within the $k_{T}$-factorization 
approach discussed in the previous section. We have written a new code called 
\pptoll based on the $k_{T}$-factorization formalism which performs all the 
integrations with the help of VEGAS method \cite{vegas} and calculates many 
differential distributions, both of single-particle type as well as of 
correlation type. The code used by us is similar to the one  developed for 
production of $c\bar{c}$ quarks (e.g., see Ref.~\cite{LS2006}).

Let us start the presentation of our results for integrated cross sections. 
In Table~\ref{table:cross_sections} we present integrated cross sections for 
two different sets of cuts for single muons:

They will be called \emph{low $p_T$} ($p_T >$ 3 GeV) and \emph{high $p_T$} ($p_T >$
15 GeV) in the following for brevity. 
We have collected results with different deep-inelastic structure functions 
\cite{SU2000,SU2000_GSR,Fiore:2002re}. While for the low-$p_T$ cuts the elastic-elastic 
contribution dominates, for the high-$p_T$ cuts all contributions are of the 
similar size. The Szczurek-Uleshchenko (SU) $F_{2}$ structure function~\cite{SU2000} gives larger 
cross section than, e.g., the parametrization obtained by Fiore et al. \cite{Fiore:2002re} 
of the CLAS collaboration data.
The numbers presented in Table~\ref{table:cross_sections} obtained with the 
code \pptoll agrees with the Monte Carlo code \lpair \cite{lpair}. We 
note large differences between results obtained with the SU 
 and the parametrizations by Fiore et al (labeled as FFJLM) \cite{Fiore:2002re} for 
inelastic-inelastic contribution for different models of $F_{2}$ structure 
functions, especially for high-$p_T$ cuts. We shall return to this point in the 
conclusion section.

\begin{figure}[!h]
  \includegraphics[width=.52\textwidth]{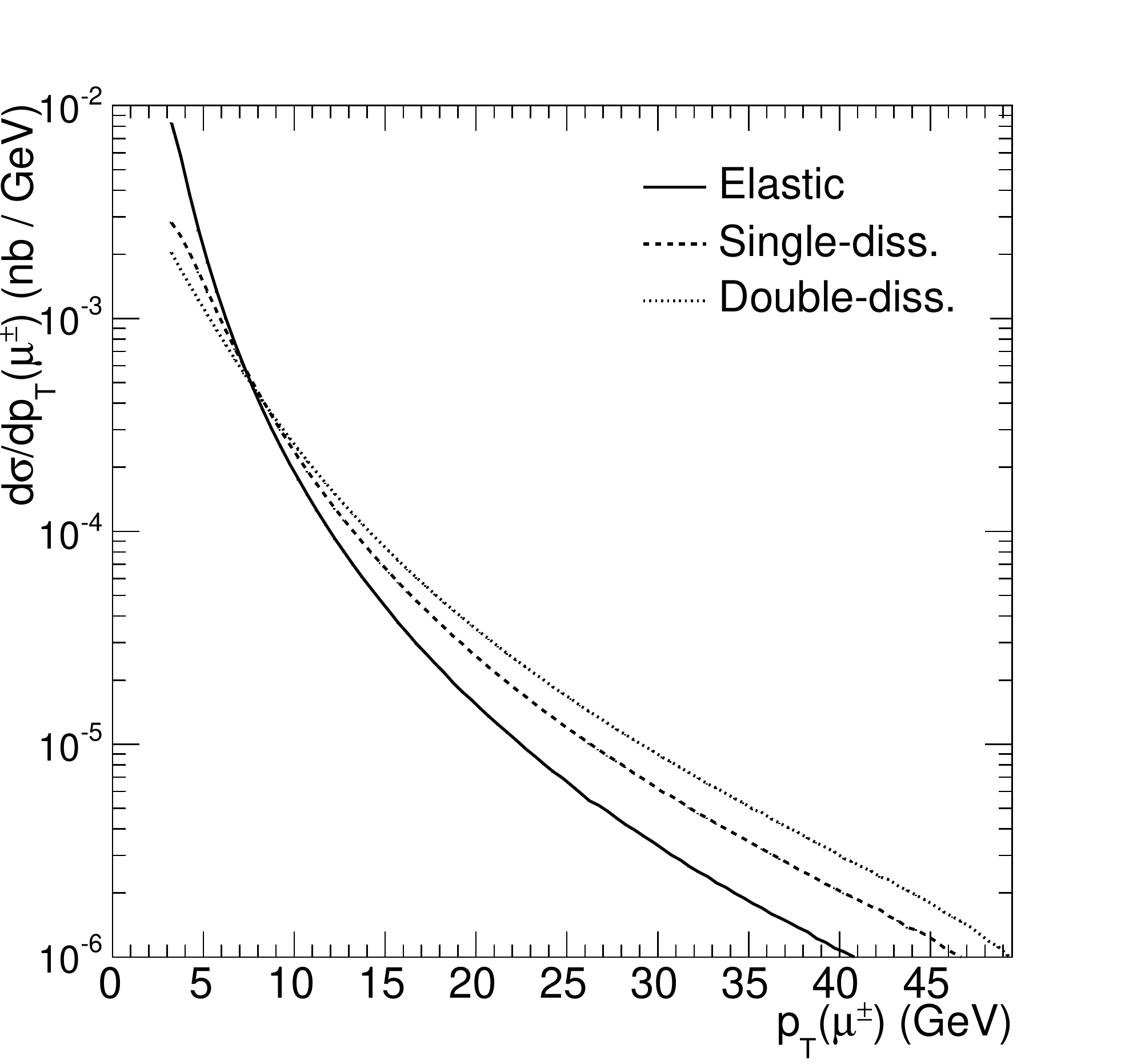}
  \includegraphics[width=.52\textwidth]{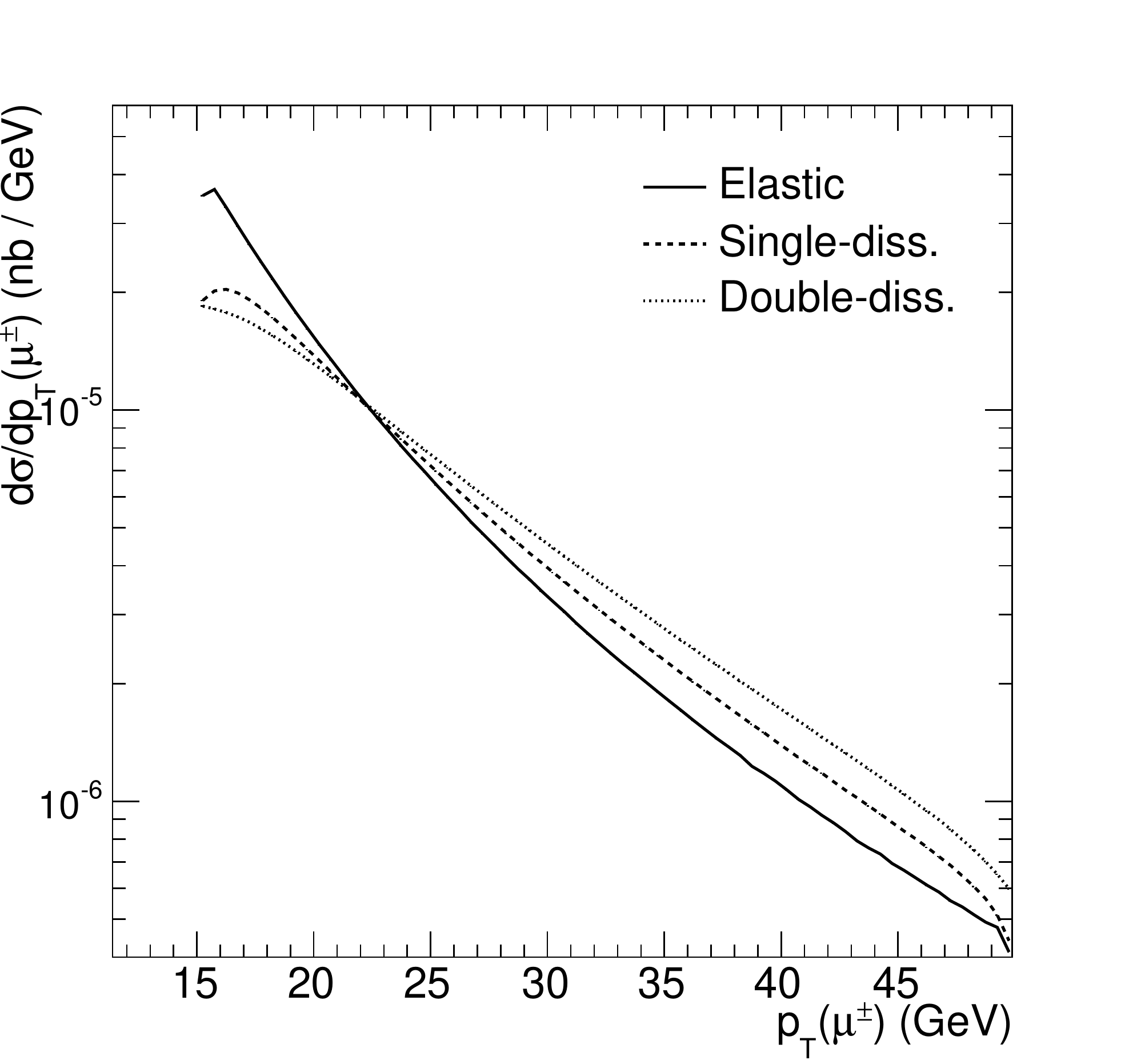}
  \caption{\label{fig:dsig_dpt}
  Muon transverse momentum distributions for low-$p_T$ (left) and 
  high-$p_T$ (right) cuts. The different lines represent elastic-elastic
  (solid), elastic-inelastic (dashed), inelastic-elastic (dotted) and 
  inelastic-inelastic (dash-dotted) mechanism at $\sqrt{s}$~=~7~TeV.
  The additional cuts of $\pt(\mu^{\pm})<$~50~GeV, 
  $|\eta(\mu^{\pm})|<$~2.5, and $M_{X}=$~[1.07,1000.00]~GeV are used in all
  cases.
  }
\end{figure}

\begin{center}
\begin{table}
  \centering
  \begin{tabular}{|l|cc|cc|}
  \hline
  {\bf Contribution} & \multicolumn{2}{c|}{\bf Low $p_T$ ($p_T >$~3~GeV)} &
  \multicolumn{2}{c|}{\bf High $p_T$ ($p_T >$~15~GeV)} \\
  \hline
                      & SU     & FFJLM  & SU    & FFJLM                        \\
  \hline
  elastic-elastic     &  16.39 &   --   &  0.297 &  --                         \\
  elastic-inelastic   &  10.82 &  10.28 &  0.300 &  0.302                      \\
  inelastic-elastic   &  10.82 &  10.28 &  0.300 &  0.302                      \\
  inelastic-inelastic &   9.25 &   6.39 &  0.329 &  0.301                      \\
  \hline
  \end{tabular}
  \caption{\label{table:cross_sections}
  Cross section (in pb) for $\mu^{+}\mu^{-}$ production for two different sets of 
  $\pt$ cuts and two different structure function models at 
  $\sqrt{s}$~=~7~TeV. The additional cuts of $|\eta(\mu^{\pm})|<$~2.5,
  and $M_{X}=$~[1.07,1000.00]~GeV are used in all cases.}
\end{table}
\end{center}
In Figure~\ref{fig:dsig_dpt} we present distributions in transverse momentum 
of muons. The distributions drop quickly with growing muon transverse momentum. 
Different contributions have fairly similar shapes. This reflects matrix element 
dependence on muon transverse momentum which is the same for different 
components.

\begin{figure}[!h]
  \includegraphics[width=.52\textwidth]{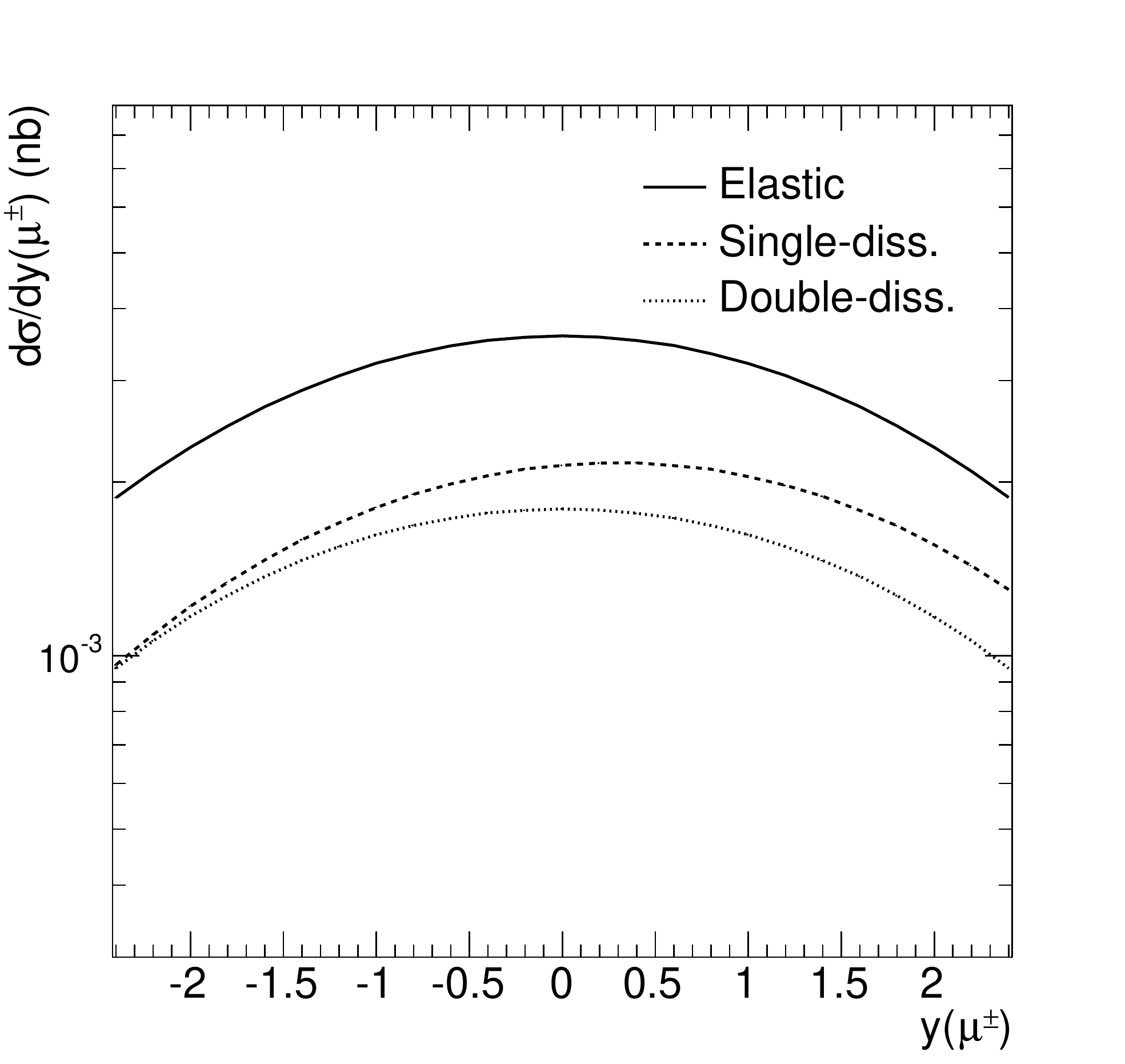}
  \includegraphics[width=.52\textwidth]{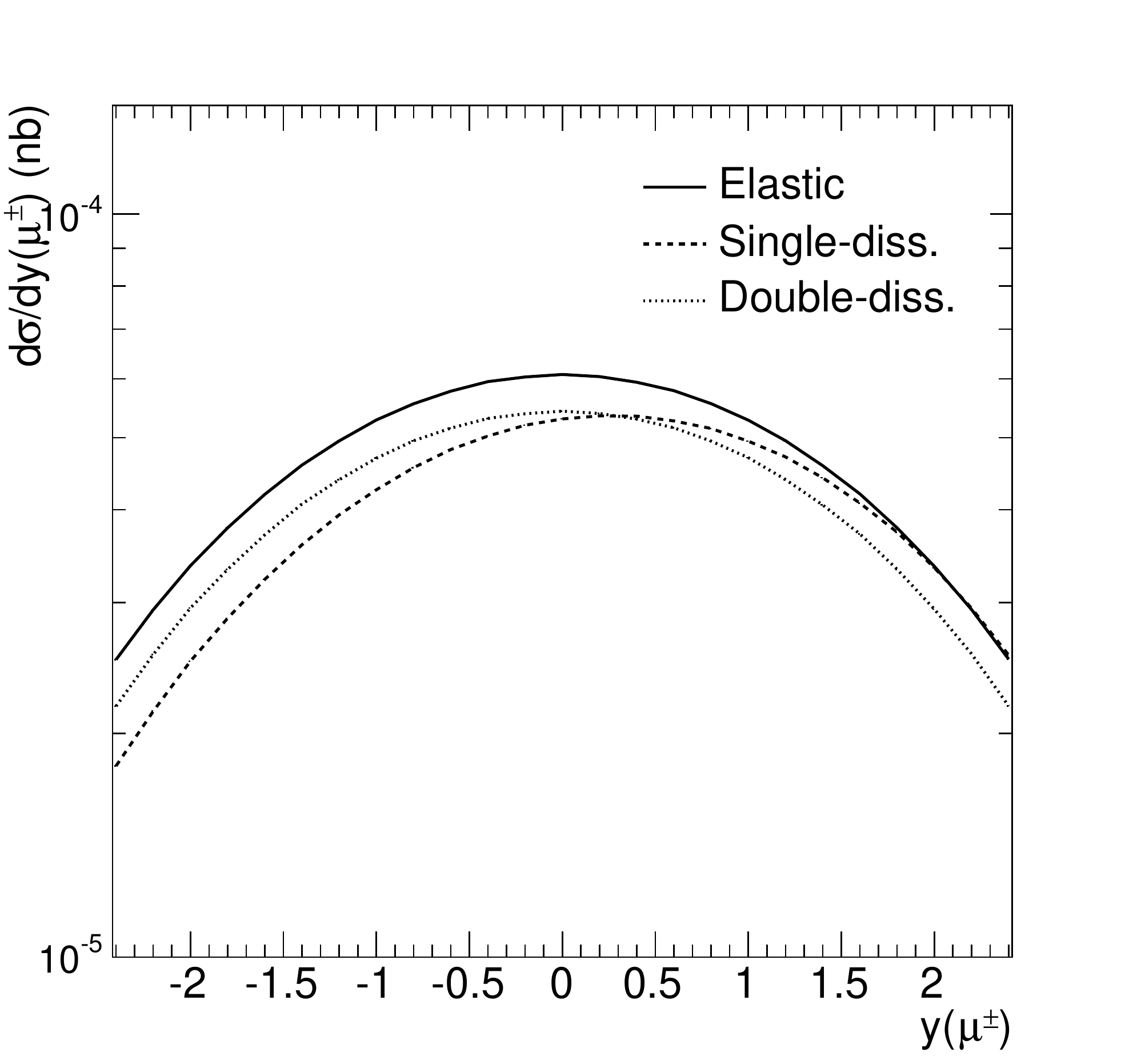}
  \caption{\label{fig:dsig_dy}
  Muon rapidity distributions for low-$p_T$ (left) and high-$p_T$ (right) cuts at 
  $\sqrt{s}$~=~7~TeV. The meaning of the curves is the same as in 
  Figure~\ref{fig:dsig_dpt}. The additional cuts of $\pt(\mu^{\pm})<$~50~GeV, 
  $|\eta(\mu^{\pm})|<$~2.5, and $M_{X}=$~[1.07,1000.00]~GeV are used in all
  cases.
  }
\end{figure}

Figure~\ref{fig:dsig_dy} presents the corresponding distributions in muon 
rapidity. While elastic-elastic and inelastic-inelastic contributions are 
symmetric with respect to $y$~=~0, the mixed contributions have maxima at $y>$~0 
(elastic-inelastic) or $y<$~0 (inelastic-elastic).

\begin{figure}[!h]
  \includegraphics[width=.52\textwidth]{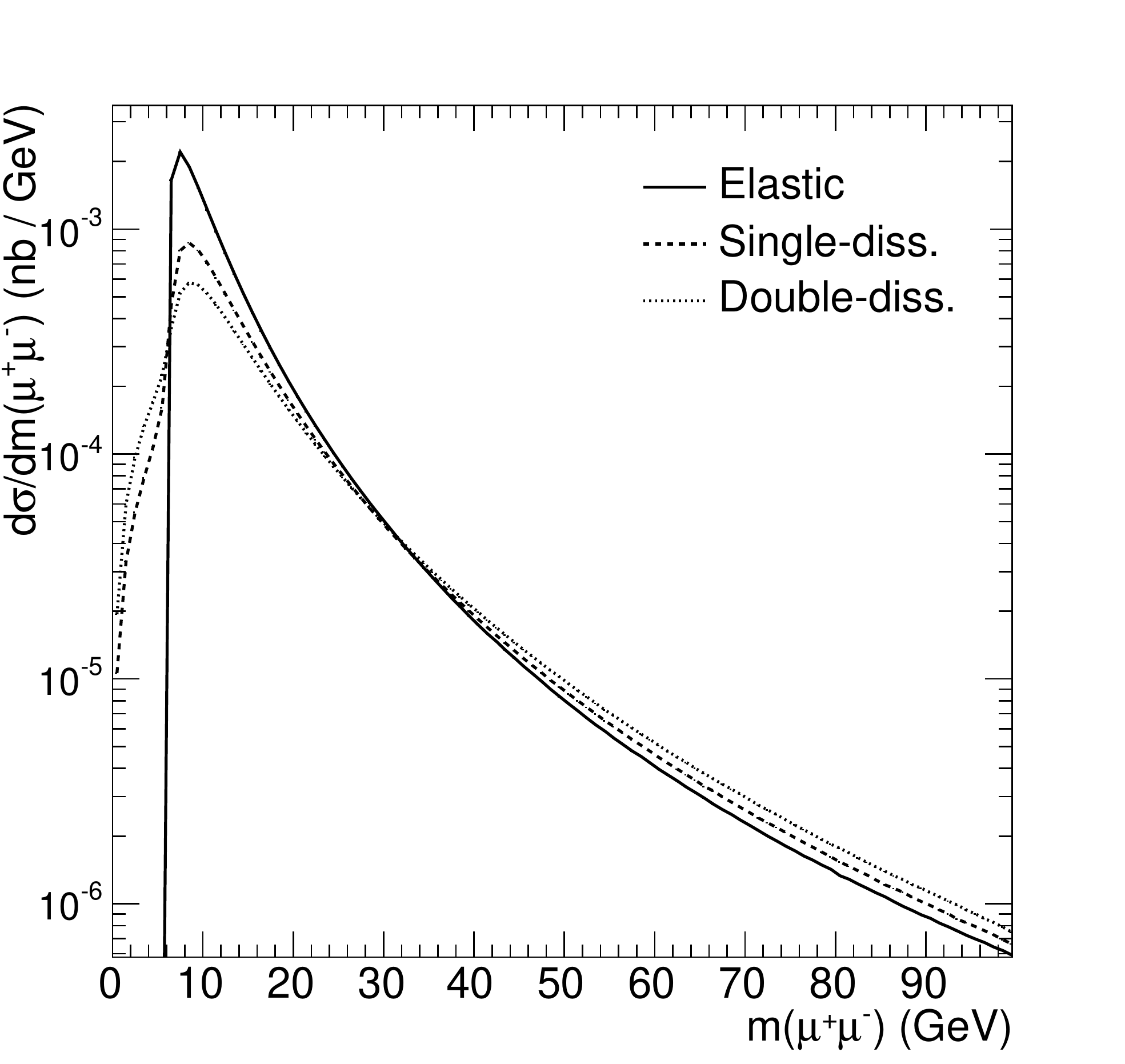}
  \includegraphics[width=.52\textwidth]{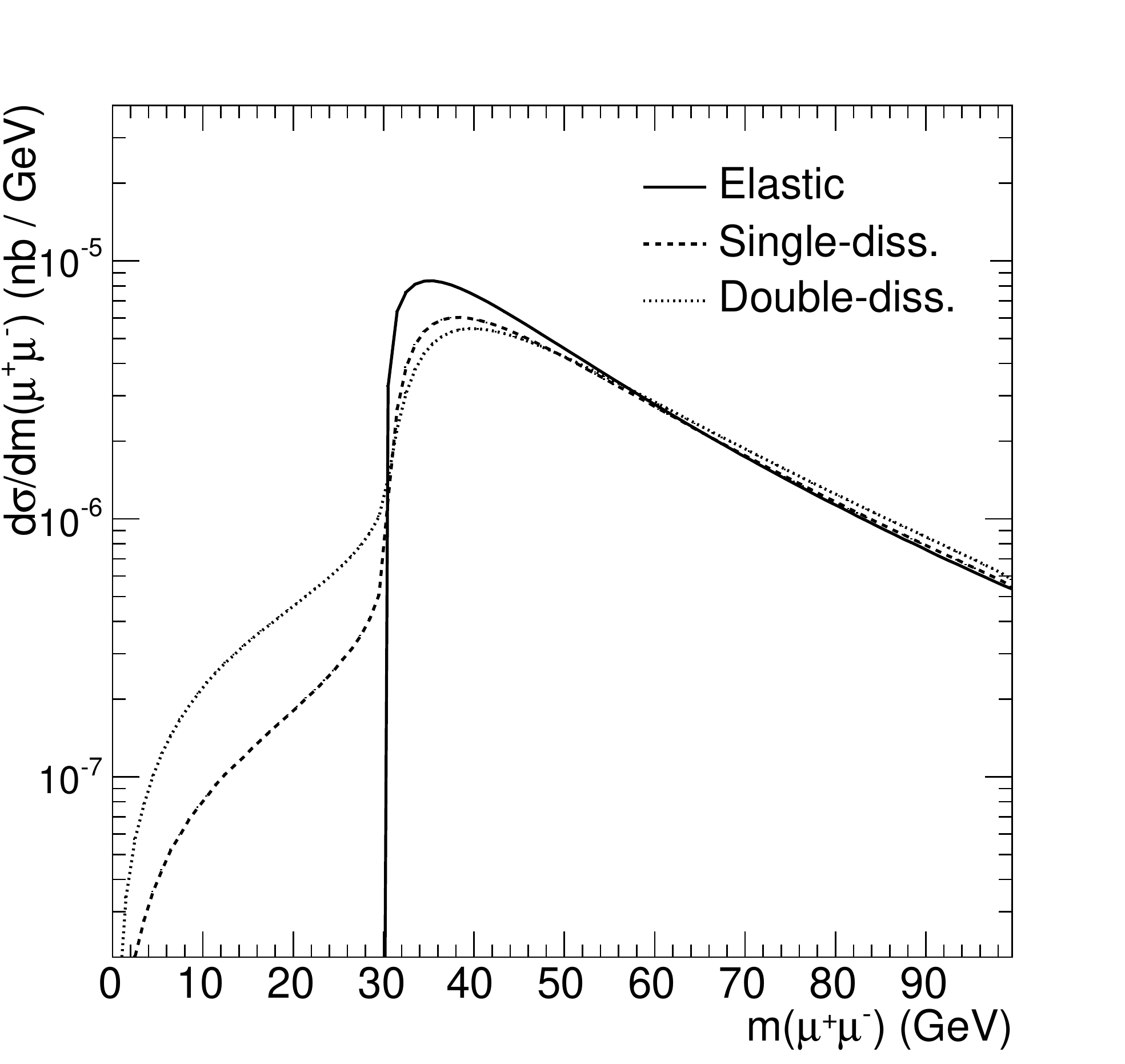}
  \caption{\label{fig:dsig_dMll}
  Muon pair invariant mass distributions for low-$p_T$ (left) and high-$p_T$ (right) 
  cuts at $\sqrt{s}$~=~7~TeV. The meaning of the curves is the same as 
  in Figure~\ref{fig:dsig_dpt}. The additional cuts of $\pt(\mu^{\pm})<$~50~GeV, 
  $|\eta(\mu^{\pm})|<$~2.5, and $M_{X}=$~[1.07,1000.00]~GeV are used in all
  cases.
  }
\end{figure}

Now we go to correlation observables which are particularly interesting 
and can be studied conveniently in our formalism. The invariant 
mass distribution shown in Figure~\ref{fig:dsig_dMll} is a first
example. All different contributions have similar shapes.

\begin{figure}[!h]
  \includegraphics[width=.52\textwidth]{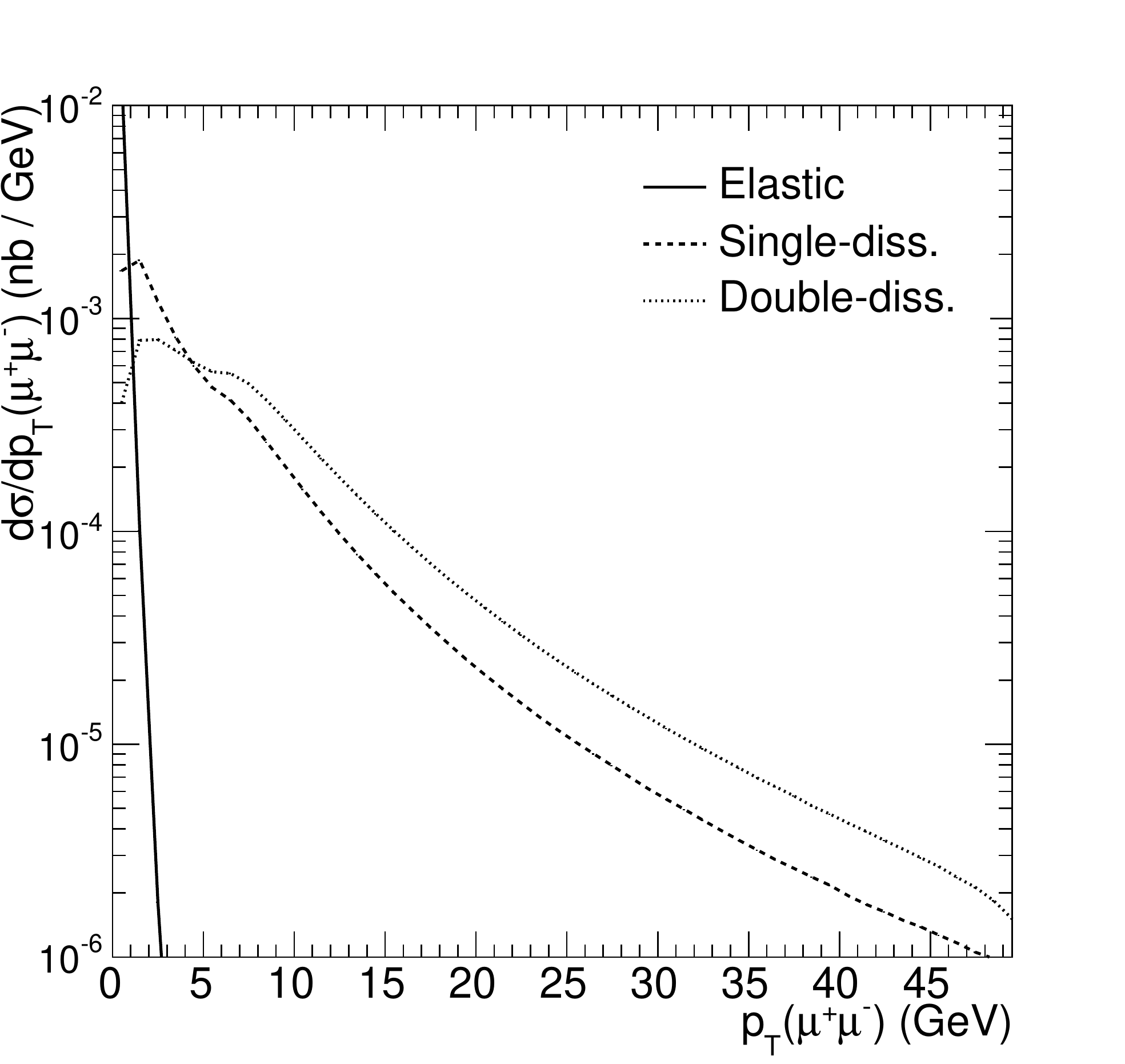}
  \includegraphics[width=.52\textwidth]{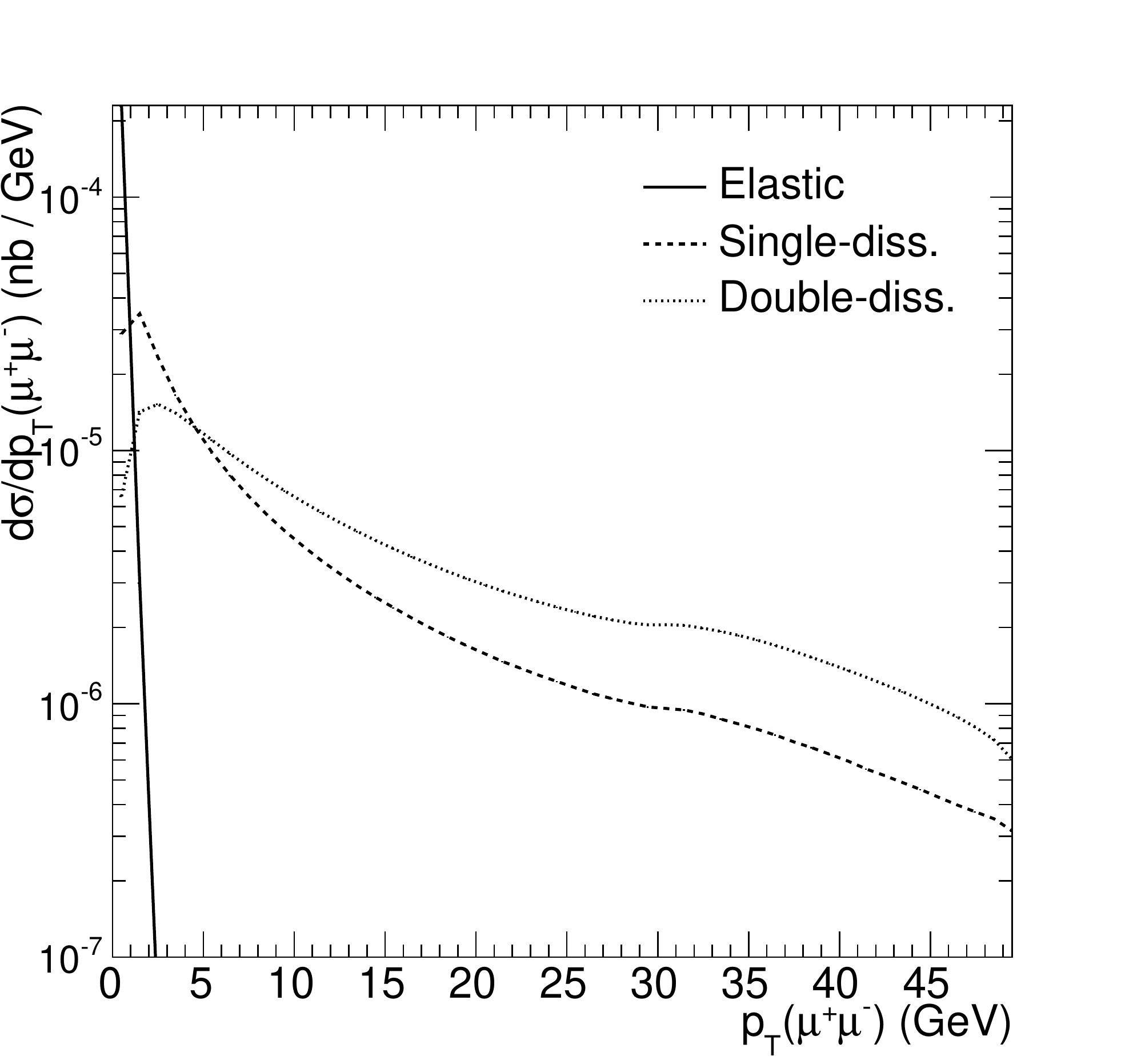}
  \caption{\label{fig:dsig_dptsum}
  Muon pair transverse momentum distributions for low-$p_T$ (left) and 
  high-$p_T$ (right) cuts at $\sqrt{s}$~=~7~TeV. The meaning of 
  the curves is the same as in Figure~\ref{fig:dsig_dpt}. The additional cuts of $\pt(\mu^{\pm})<$~50~GeV, 
  $|\eta(\mu^{\pm})|<$~2.5, and $M_{X}=$~[1.07,1000.00]~GeV are used in all
  cases.
  }
\end{figure}

In Figure~\ref{fig:dsig_dptsum} we show distribution in transverse momentum of 
the muon pair. The elastic-elastic contribution is peaked sharply at 
$p_{T\textrm{sum}}\approx$~0, while other contributions have long tails towards large 
$p_{T\textrm{sum}}$. In the collinear approximation used, e.g., in Ref.~\cite{MSS2011}, the 
transverse momenta of two muons are fully balanced and this type 
of distributions cannot be studied in that approximation.

\begin{figure}[!h]
  \includegraphics[width=.52\textwidth]{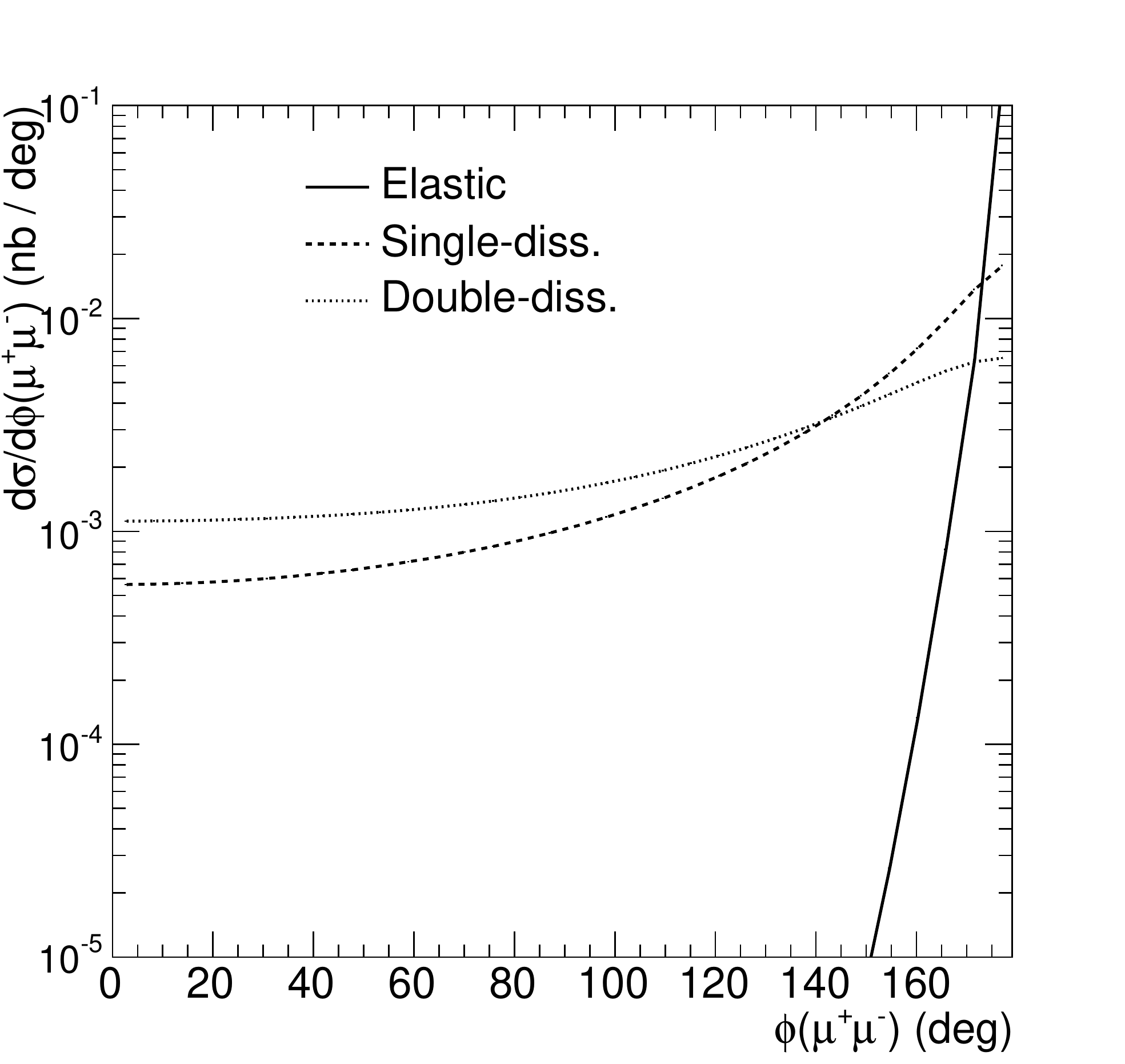}
  \includegraphics[width=.52\textwidth]{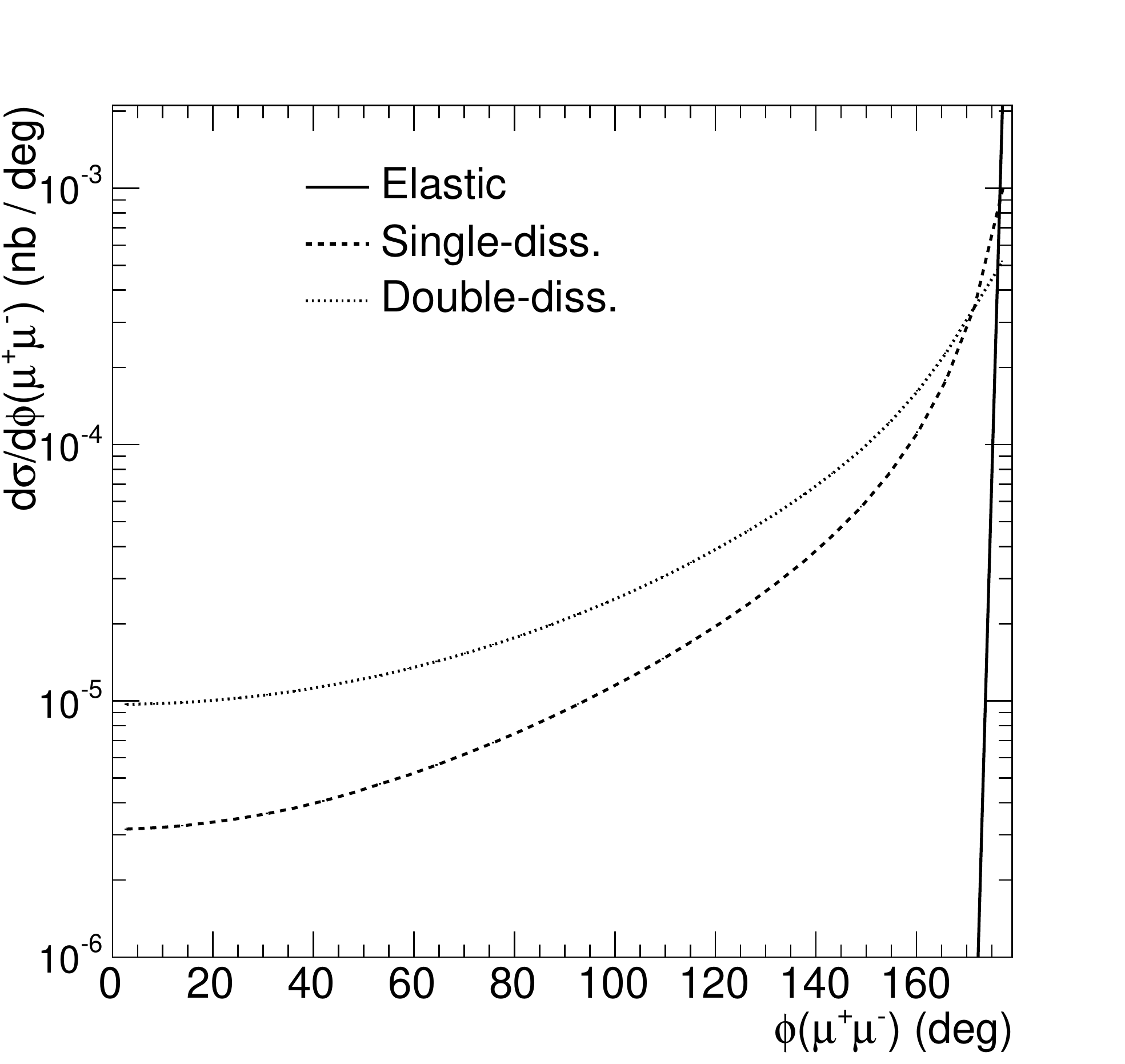}
  \caption{\label{fig:dsig_dphid}
  Distribution in azimuthal angle between muons for low-$p_T$ (left) and high-$p_T$ 
  (right) cuts at $\sqrt{s}$~=~7~TeV. The meaning of the curves is the same as 
  in Figure~\ref{fig:dsig_dpt}. The additional cuts of $\pt(\mu^{\pm})<$~50~GeV, 
  $|\eta(\mu^{\pm})|<$~2.5, and $M_{X}=$~[1.07,1000.00]~GeV are used in all
  cases.
  }
\end{figure}

Finally, Figure~\ref{fig:dsig_dphid} shows the azimuthal correlations between 
outgoing $\mu^+$ and $\mu^-$. The elastic-elastic contribution is sharply peaked 
at $\Delta \phi \approx \pi$ while two other distributions have long tails down 
to $\Delta \phi = 0$. 

Different models of structure functions have been proposed in the literature 
for $F_{2}$ structure function in broad range of $x$ and $Q^{2}$. In different 
phase space regions, defined mostly by experimental cuts, one tests $F_{2}$ in 
different ranges of its arguments.

Let us concentrate now on the region of large transverse momentum of the pair. 
In this region, as discussed above, the elastic-elastic contribution gives 
negligible contribution and inelastic contributions have to be considered.

\begin{figure}[!h]
  \includegraphics[width=.52\textwidth]{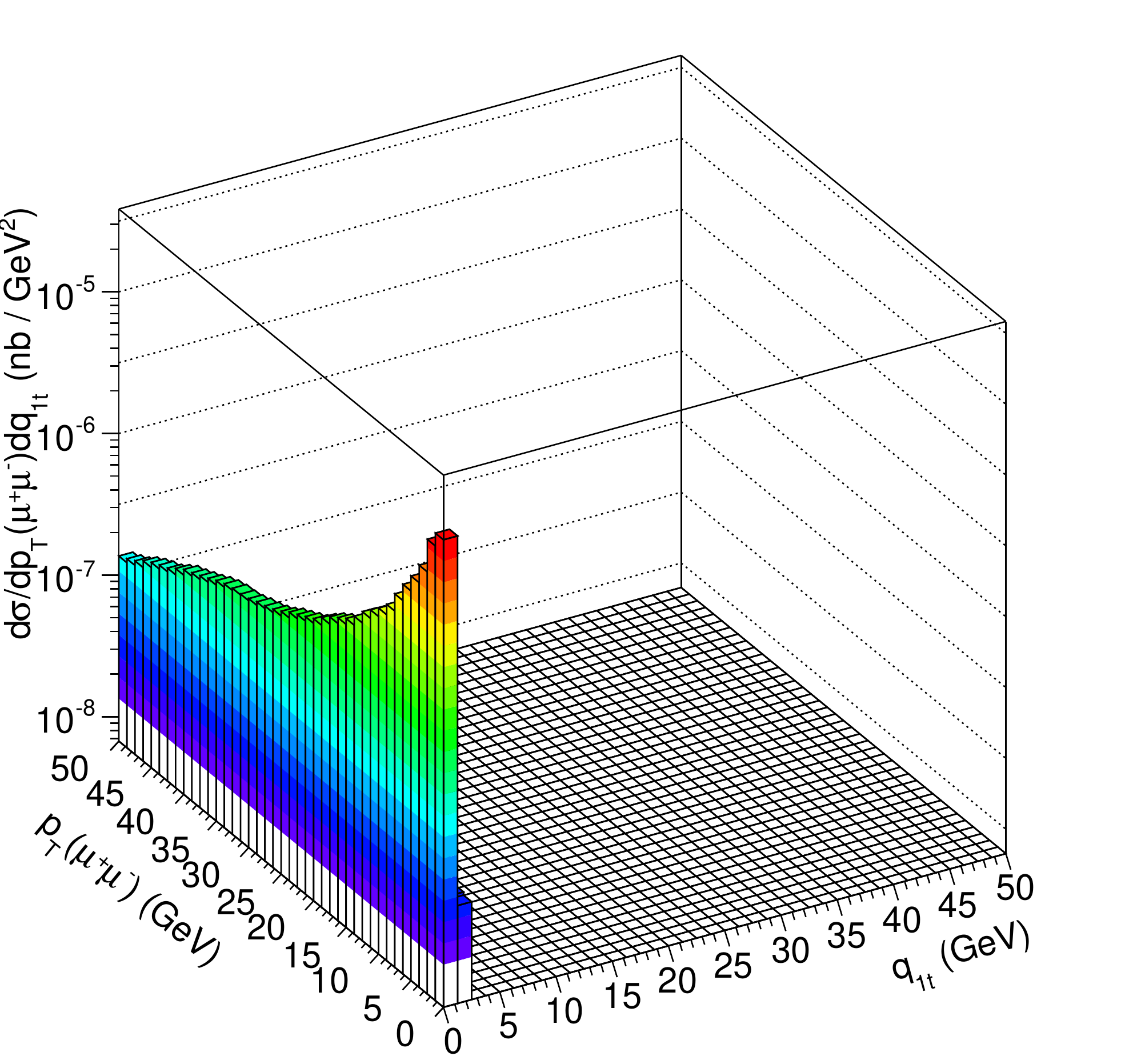}
  \includegraphics[width=.52\textwidth]{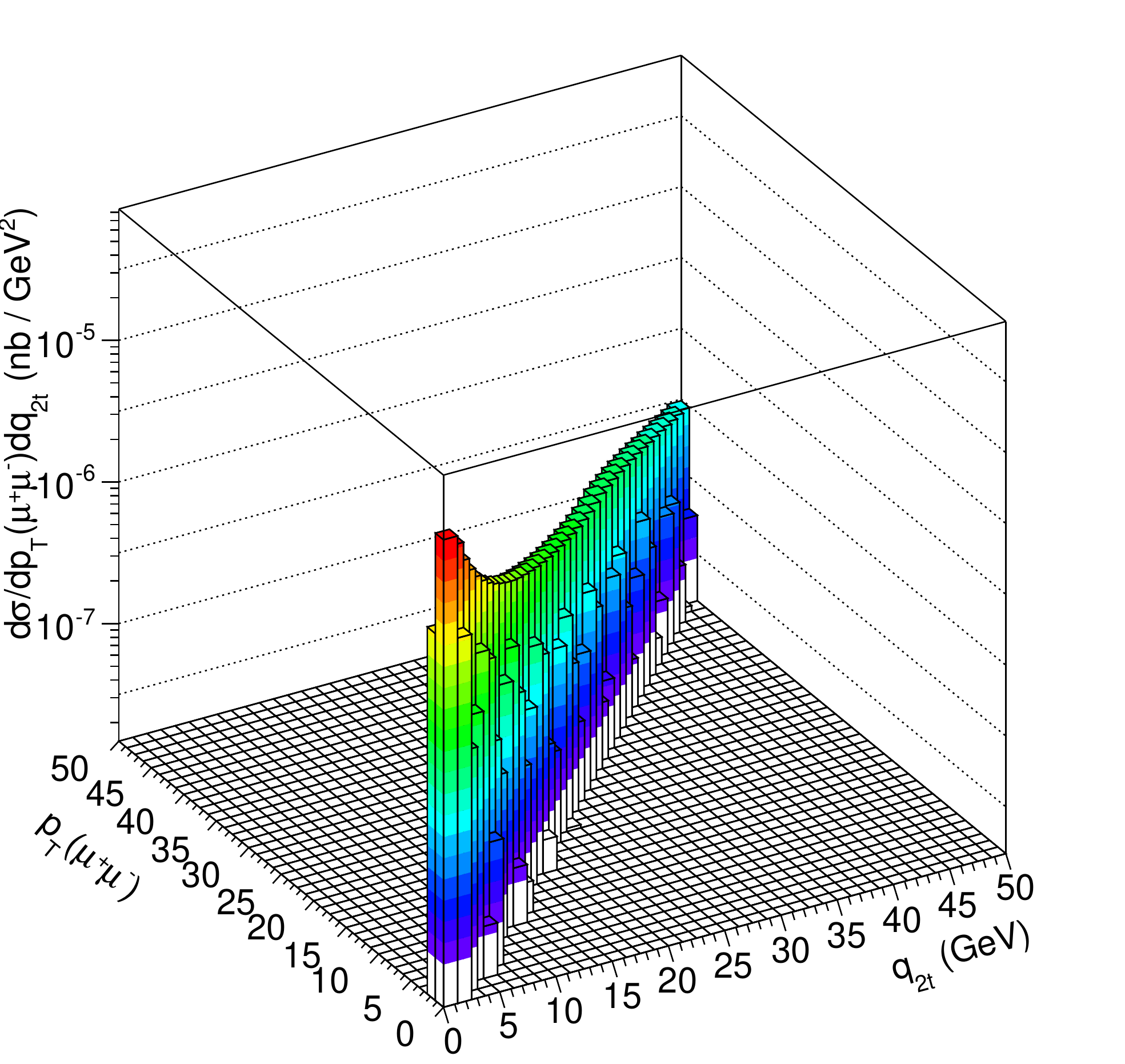}
  \includegraphics[width=.52\textwidth]{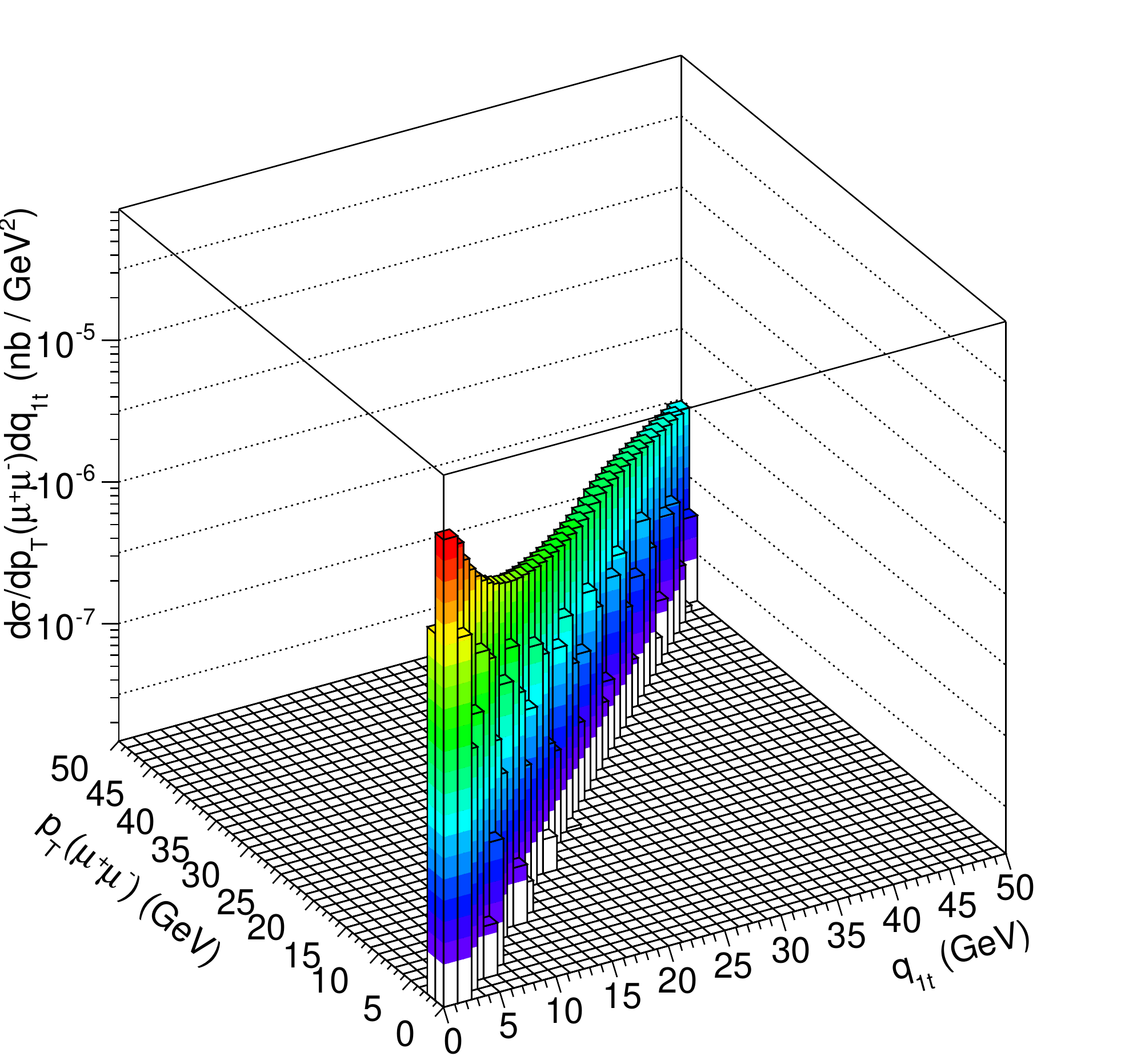}
  \includegraphics[width=.52\textwidth]{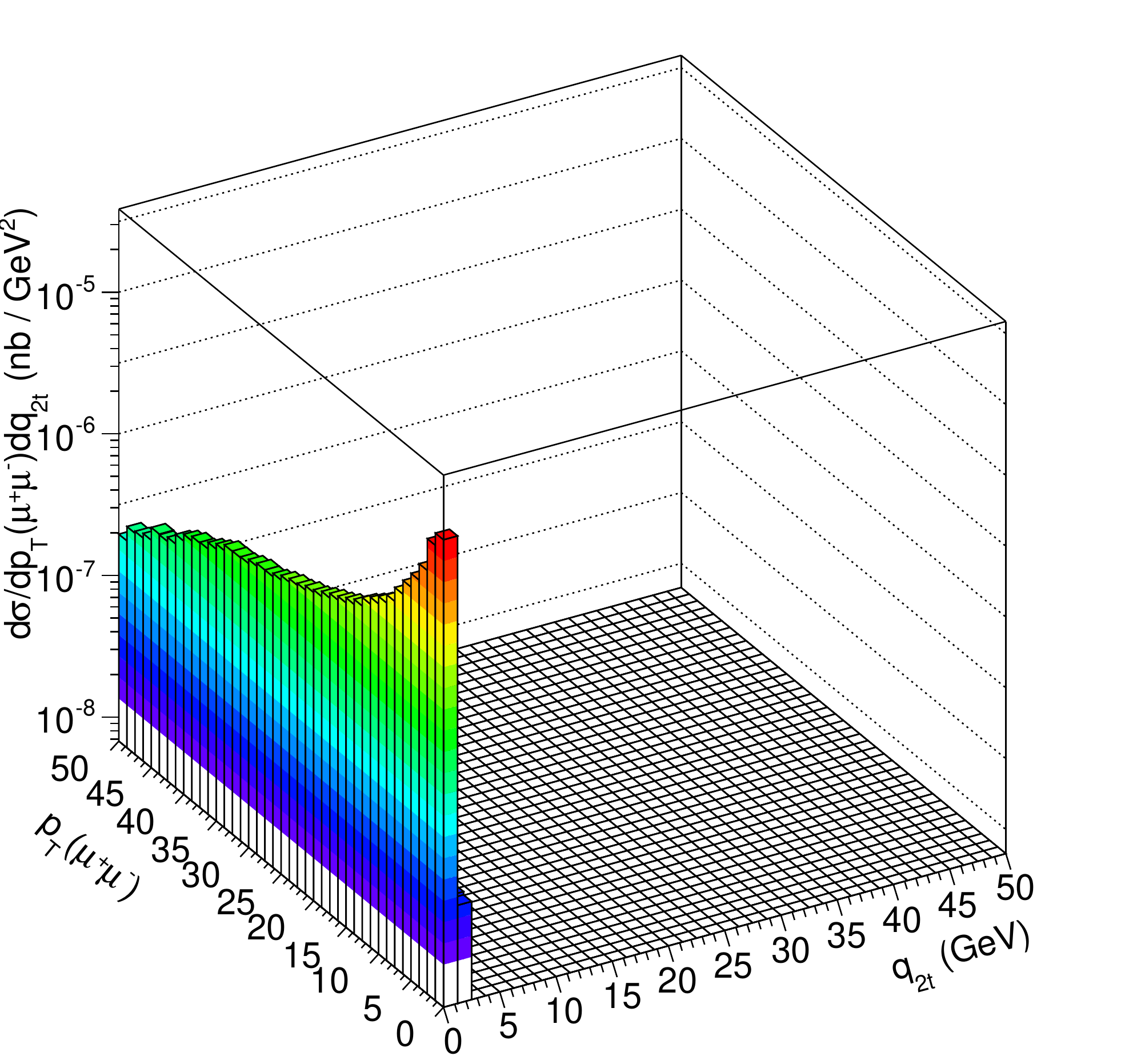}
  \caption{\label{fig:map_ptsumqit_mixed}
  Two-dimensional distribution in $p_{T\textrm{sum}}$ and one of transverse 
  momentum ($q_{1T}$ or $q_{2T}$) of the virtual photon for elastic-inelastic 
  (top row) and inelastic-elastic (bottom row) components at 
  $\sqrt{s}$~=~7~TeV. The additional cuts of $\pt(\mu^{\pm})<$~50~GeV, 
  $|\eta(\mu^{\pm})|<$~2.5, and $M_{X}=$~[1.07,1000.00]~GeV are used.
  }
\end{figure}

We first consider the mixed components (elastic-inelastic or inelastic-elastic). 
In Figure~\ref{fig:map_ptsumqit_mixed} we show two-dimensional distributions 
in transverse momentum of the muon pair ($p_{T\textrm{sum}}$) and transverse momentum 
of one of the virtual photons. We observe a strong correlation between 
$p_{T\textrm{sum}}$ and transverse momentum of the ``inelastic photon'' (the photon 
which leaves proton in the excited state). For this photon the virtuality is
$Q_i^2 \approx -q_{i,t}^2$. Large transverse momentum means therefore also 
large virtuality. For large virtualities partonic structure functions which 
undergo QCD DGLAP evolution are adequate. The SU structure function, 
which in this region is just partonic $F_{2}$, is therefore a better choice 
than the Suri-Yennie (SY) parametrization~\cite{suri} used in the \lpair event 
generator \cite{lpair}, which is adequate for small $Q^{2}$ and does not 
guarantee correct dependence at large $Q^{2}$. Then, standard well-known 
partonic structure functions should be rather used.

\begin{figure}[!h]
  \includegraphics[width=.52\textwidth]{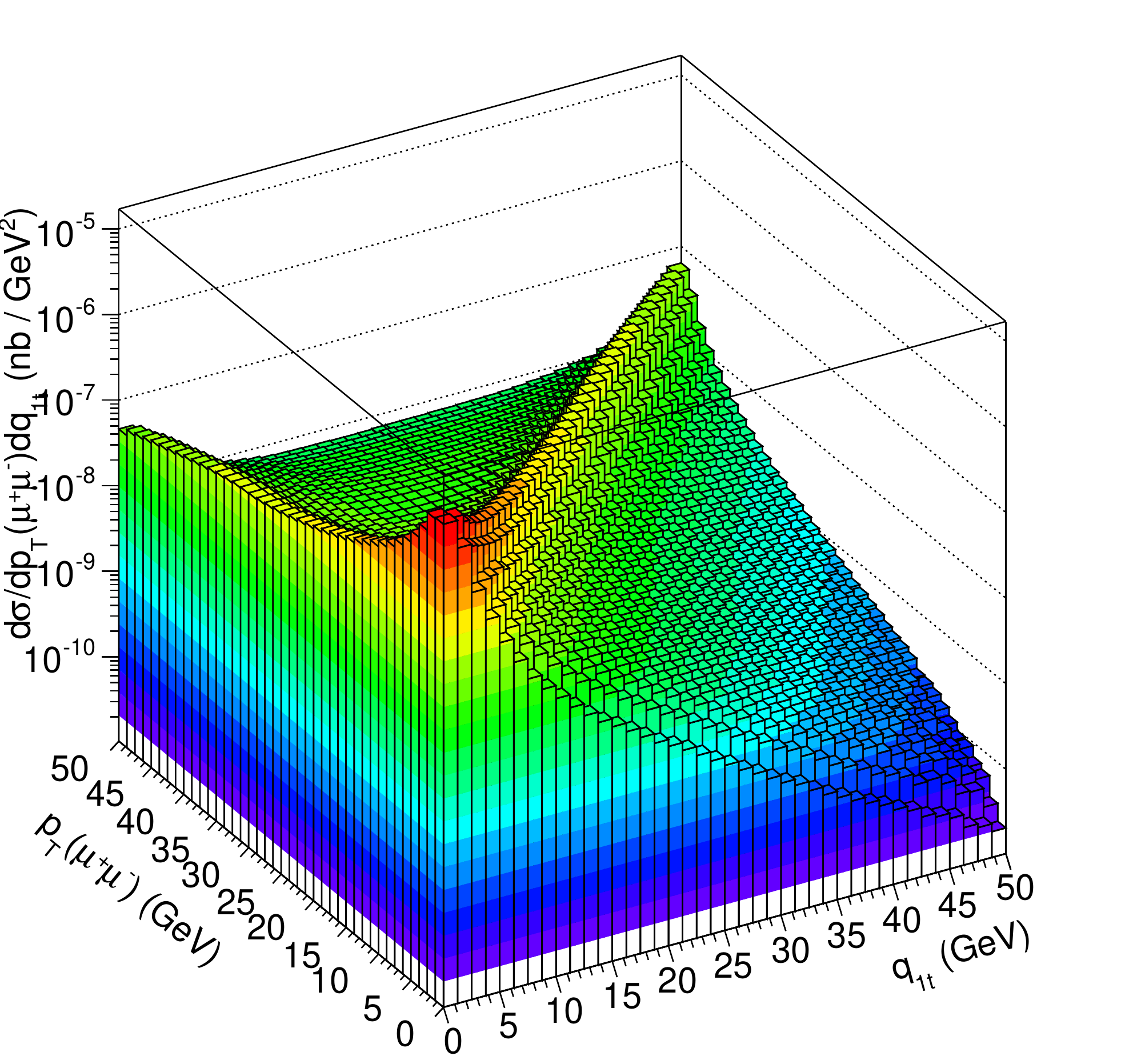}
  \includegraphics[width=.52\textwidth]{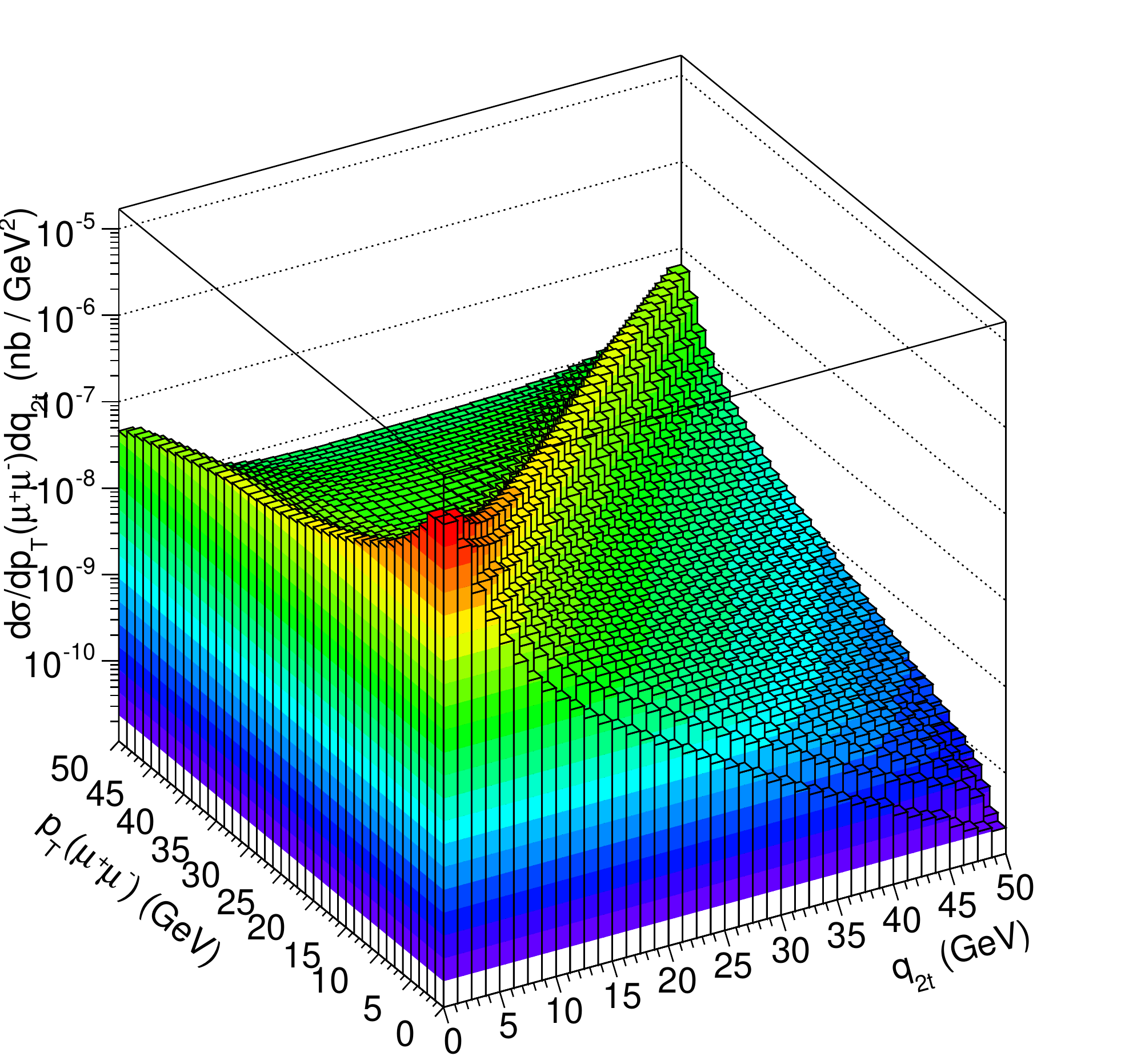}
  \caption{\label{fig:map_ptsumqit_ineine}
  Two-dimensional distribution in $p_{T\textrm{sum}}$ and one of transverse
  momentum of the virtual photon for inelastic-inelastic component at 
  $\sqrt{s}$~=~7~TeV for $p_{1T}, p_{2T} >$~15~GeV. The additional cuts of $\pt(\mu^{\pm})<$~50~GeV, 
  $|\eta(\mu^{\pm})|<$~2.5, and $M_{X}=$~[1.07,1000.00]~GeV are used.
  }
\end{figure}

For inelastic-inelastic component the situation is, however, more complicated 
as is shown in Figure~\ref{fig:map_ptsumqit_ineine}. One can observe two 
characteristic ridges around the diagonal $p_{T\textrm{sum}} = q_{it}$ and for small 
$q_{1T}\approx$~0 or $q_{2T}\approx$~0. 

\begin{figure}[!h]
  \centering
  \includegraphics[width=.52\textwidth]{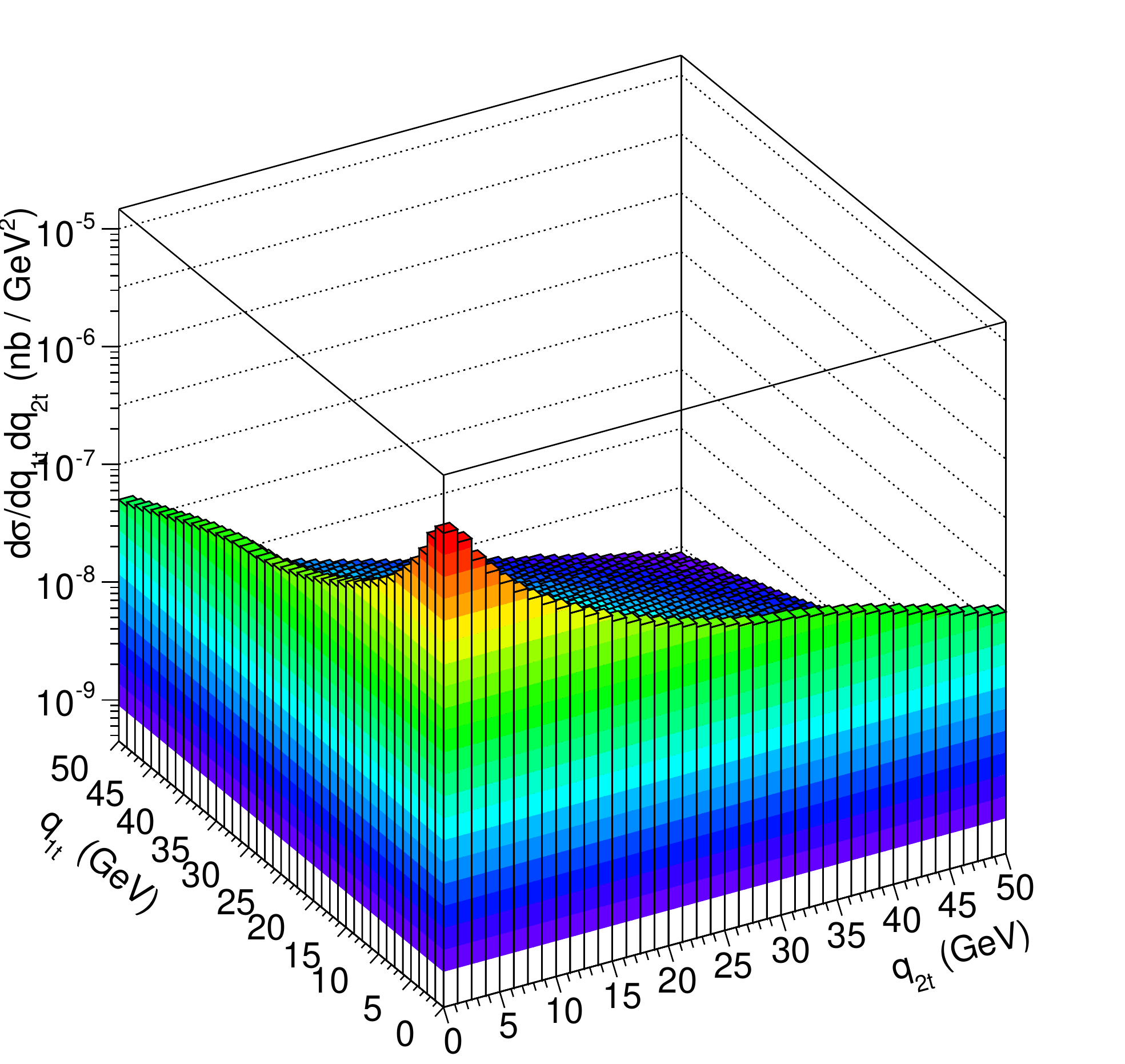}
  \caption{\label{fig:map_q1tq2t_ineine}
  Two-dimensional distribution in ($q_{1T}$,$q_{2T}$) for 
  inelastic-inelastic component at $\sqrt{s}$~=~7~TeV and for 
  $p_{1T}, p_{2T} >$~15~GeV. The additional cuts of $\pt(\mu^{\pm})<$~50~GeV, 
  $|\eta(\mu^{\pm})|<$~2.5, and $M_{X}=$~[1.07,1000.00]~GeV are used.
  }
\end{figure}

To understand the two-ridge structures of Fig.~\ref{fig:map_ptsumqit_ineine}, 
in Fig.~\ref{fig:map_q1tq2t_ineine} we show a two-dimensional map in 
($q_{1T},q_{2T}$). The corresponding cross section is peaked along $q_{1T}$ 
and $q_{2T}$ axes, where in tails $q_{1T}$ is small and $q_{2T}$ is large or
$q_{1T}$ is large and $q_{2T}$ is small. One clearly sees that for 
inelastic-inelastic contribution, even at large $p_{T\textrm{sum}}$, unlike for mixed 
components one is sensitive to low virtuality physics of resonance production. 
Therefore, a careful treatment of the low-virtuality region of $F_{2}$ is for 
inelastic-inelastic component particularly important. Further studies are 
clearly needed.

\section{LPAIR Monte Carlo studies}
\label{sec:mc-studies}

\subsection{LPAIR event generator}
\label{lpair}

We performed an additional study using the \lpair Monte Carlo (MC) event generator, 
which accounts for the electromagnetic production of lepton pairs in $e^{+}e^{-}$, 
$ep$ and $pp$\footnote{In fact, the second colliding proton is treated in 
\lpair only elastically, since there is no calculation for the dissociation of both 
protons in original version of the code.} collisions~\cite{lpair,vermaseren}. 
One should note that the \lpair generator does not take into account hadronic corrections to the
cross sections, due to rescattering, for example. 
\lpair has been extensively used by experiments at HERA, Tevatron and the LHC.

Instead of employing the equivalent photon approximation (EPA), \lpair accounts 
for the two-photon interaction in a $2\to 4$ process by computing 
the full matrix element, e.g., for the $pp\to p\ell^{+}\ell^{-}p$
diagram~\cite{vermaseren}. The hadronic structure of the proton is then effectively taken into
account by multiplying the proton charge by appropriate form factors 
or structure functions. 

In the case of an exclusive production (Fig.~\ref{diagrams}a), the probability 
of a photon exchange from the proton is accounted by the electromagnetic form 
factor of the proton in the dipole approximation:

\beq
  F_{p}(Q^{2}) = \frac{G_{E}(Q^{2}) + kG_{M}(Q^{2})}{(1+k)},
\eeq
where $k = Q^{2}/4 m_{p}^{2}$ and
\beq
  G_{E}(Q^{2}) &=& \frac{1}{(1+Q^{2}/0.71\textrm{ GeV}^{2})^{2}}, \\
  G_{M}(Q^{2}) &=& \mu G_{E}(Q^{2}),
\eeq
where $G_{E}$ is the electric form factor and $G_{M}$ the magnetic form factor 
of the proton, with $\mu$~=~2.79 being the proton magnetic moment.

On the other hand, in case of semi-exclusive production (Fig.~\ref{diagrams}b,c) 
the photon exchange leads to the proton break-up. Here, the SY 
parametrization is obtained from a fit to the available data for the total $\gamma p$ cross 
section in the resonance region~\cite{suri}. 
The data employed to obtain this parametrization corresponds to the measurements
of the cross section of photonucleon processes and inelastic electron scattering,
with center-of-mass energies up to $\sqrt{s}\sim$~20~GeV.
For the purpose of proton dissociation into a hadronic system, 
\lpair is interfaced with the {\textsc{jetset}} library of 
{{\textsc{pythia6}}~\cite{pythia} in order to compute the kinematics of the 
final-state particles. Depending on the chosen structure function, 
the low-mass system can decay into $\Delta^{+}$ and $\Delta^{++}$ 
resonances while the high-mass system can decay into a larger 
multiplicity states with $\Delta$, $\eta$, and $K$ resonances.

The results obtained with \lpair in general agree well with recent measurements of the two-photon production of dileptons 
in CMS \cite{cms-papers}. One should note, however, that the predictions
for the exclusive production are weakly dependent of the modeling 
the proton structure, and it is not the case for the semi-exclusive ones.

This paper will make use of large event samples produced for the exclusive and
semi-exclusive two-photon production of dimuons in $pp$ collisions 
at $\sqrt{s}$~=~7~TeV.

\subsection{Some LPAIR results with detector acceptance}
\label{exp}

The study of the exclusive two-photon production of dimuons with the CMS detector 
has shown a good agreement of the \lpair predictions with the data, but at the same 
time a significant disagreement was noticed for a part of the semi-exclusive 
production. As seen in Figure~6 of Ref.~\cite{fsq-12-010}, the transverse 
momentum, $p_{T\mathrm{sum}}$, distribution of the muon pairs in the high-$p_{T\mathrm{sum}}$ region is not well 
described by \lpair. Specifically, the \lpair simulations do not reproduce 
the suppression seen in the data for dimuons with  $p_{T\mathrm{sum}}$ above 10~GeV. This effect 
can potentially be connected to rescattering or absorption corrections which are not 
accounted for. In order to understand the possible source of the disagreement, we 
perform here a more detailed study of the \lpair predictions of semi-exclusive 
two-photon production of dimuons at $\sqrt{s}=$~7~TeV.

As discussed in Sec.~\ref{lpair}, the SY default parametrization of 
the proton structure function in \lpair is based on data obtained with low incident energies, 
corresponding to the photon-proton center-of-mass energy $\sqrt{s}$ between 1.11 and 18.03~GeV, 
and low photon virtualities -- approximately $Q^{2}<$~10~\gev. On the other hand, the 
data collected by the CMS Collaboration for the semi-exclusive two-photon production of 
dimuons evidently contain events involving very large photon virtualities. In particular,
it concerns the region of the high $p_{T\mathrm{sum}}$ of the lepton
pairs, where the disagreement is observed.

In order to test sensitivity of the \lpair predictions to modeling of the 
structure function, we introduced a more recent parametrization available 
in the literature, such as the SU parametrization used already in 
the $k_{T}$-factorization approach. In the context of the experimental studies, we 
restrict the phase space by similar kinematic cuts as employed in Ref.~\cite{fsq-12-010}: 
$\pt(\mu^{\pm})>$~3 or 15~GeV and $|\eta(\mu^{\pm})|<$~2.5 for each muon and an invariant mass 
of the hadronic system, $M_{X}$, between 1.07 to 1000~GeV. As a result, event samples with 
one million events are produced at $\sqrt{s}$~=~7~TeV for each of the PDF parametrizations 
with the production cross sections presented in Tab.~\ref{tab2}. 


\begin{table}
  \centering
  \begin{tabular}{|lrr|}
  \hline
  \bf Parametrization     & \multicolumn{2}{c|}{$\sigma_{pp}(\gamma\gamma\to\mu^{+}\mu^{-})$ (pb)} \\ \hline
                          & $\pt>$~3~GeV  & $\pt>$~15~GeV       \\ 
  \hline
  Elastic-Elastic                         & 16.28        & 0.29 \\
  Inelastic-Elastic, Suri-Yennie          & 11.77        & 0.33 \\
  Inelastic-Elastic, Szczurek-Uleshchenko & 9.87         & 0.30 \\
  \hline
  \end{tabular}
  \caption{\label{tab2}
  Production cross section obtained with the \lpair code for the
  kinematic cuts of $\pt(\mu^{\pm})>$~3 or 15~GeV
  and the additional cuts of $|\eta(\mu^{\pm})|<$~2.5,
  and $M_{X}=$~[1.07,1000.00]~GeV in all
  cases. 
  }
\end{table}


For the $Q^{2}$ values below the range of validity 
of the parametrizations, the approach in the SU parametrization uses a shifted value 
($Q^2+\mu^{2}$) of the factorization scale and is 
rescaled by a factor which ensures, that  $F_{2}\to 0$ when $Q^{2}\to 0$.

Figures~\ref{pts} and \ref{etas} show the $\pt$, and pseudorapidity, $\eta$, distributions of single muons 
for the SY parametrization in comparison to the SU one with $\ptmin$ of 3 and 15~GeV. 
One sees that the $\pt$ distribution have an overall agreement, while the $\eta$ distributions 
have significantly different shapes. This effect already shows that the integrated detector acceptance 
correction is sensitive to a choice of the parton densities.

Figure~\ref{phis} and \ref{pairs} show the distributions of the dimuon 
variables — the muon azimuthal angle difference and 
the pair transverse momentum. For high $\pt(\mu^{+}\mu^{-})$ the 
shapes are similar, whereas at the low $\pt$ the differences 
are more pronounced. This region is related to the lower $Q^{2}$ values,
which depends on the approach used to describe the 
photons with low virtualities and explains the shift in the peaks for each 
parametrization around $\pt\sim$~1~GeV. Similar effects are observed 
for the $\Delta\phi$ distributions. 
For completeness, in Fig.~\ref{q2plots} the distributions of the 
photon virtuality are compared. 

To better clarify the origin of these differences, in the following the results obtained with \lpair are 
compared to the complementary approach described in Section~\ref{sec:kt-results}.

\begin{figure}
  \includegraphics[width=.5\textwidth]{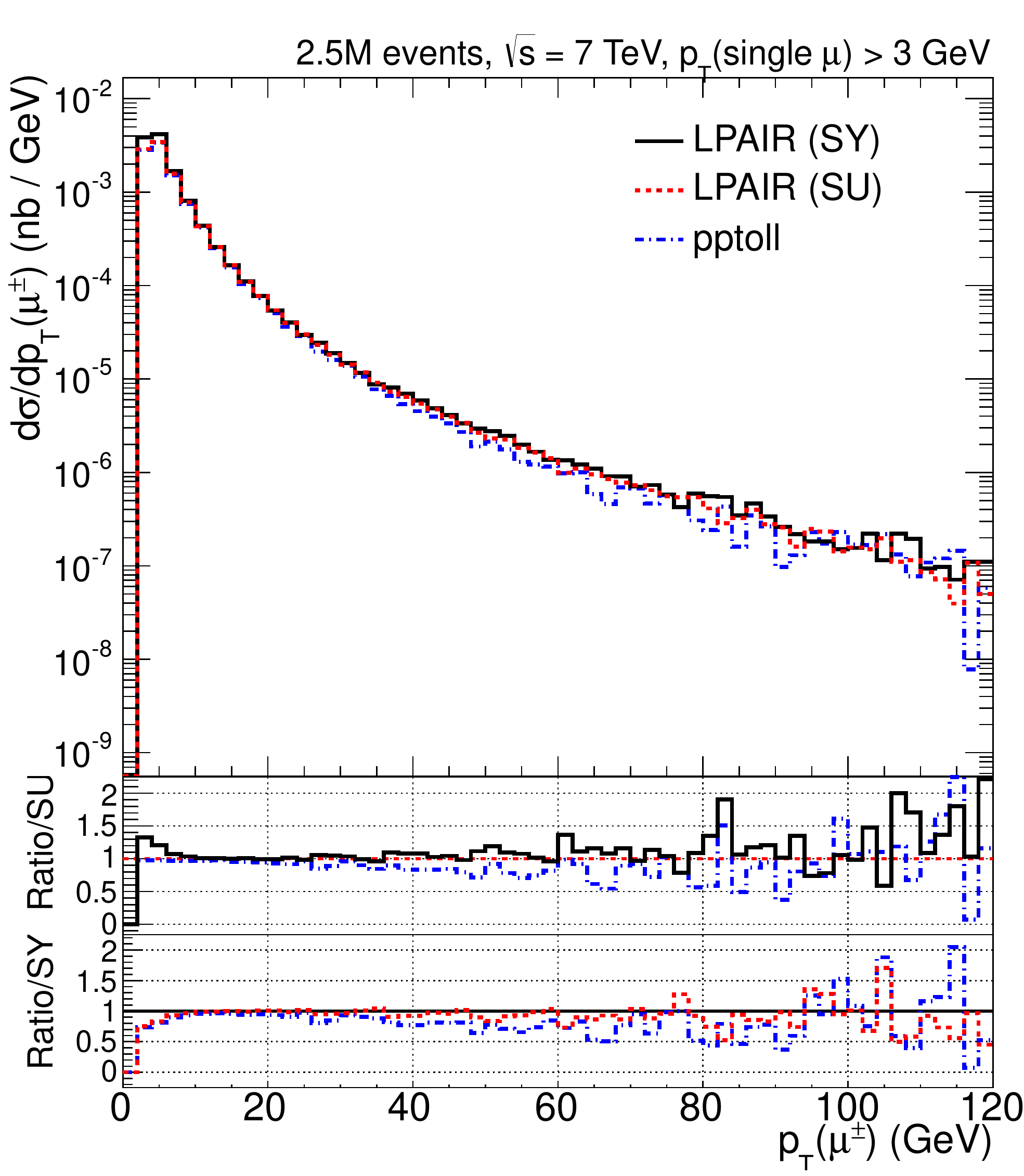}
  \includegraphics[width=.5\textwidth]{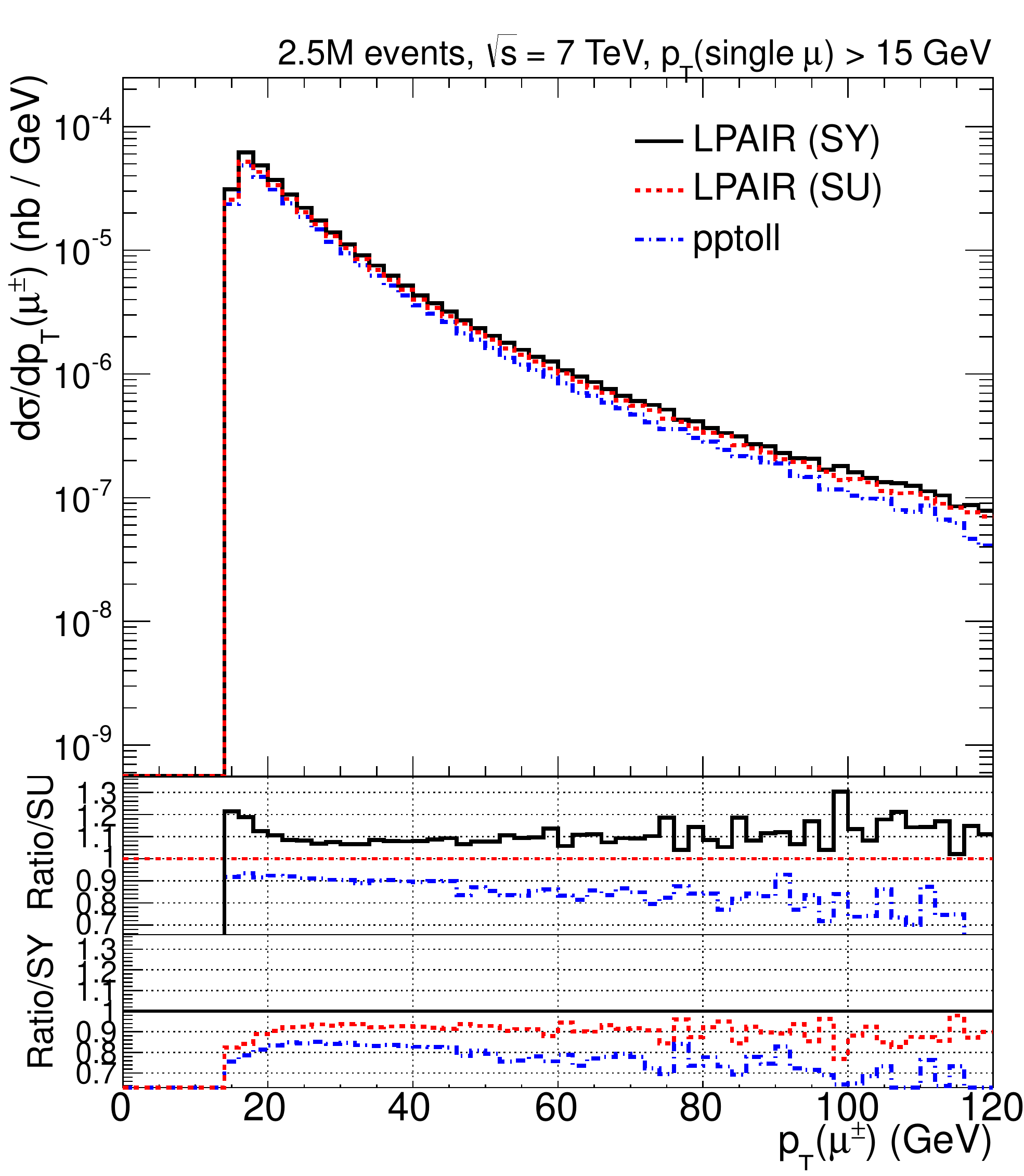}
  \caption{\label{pts}
  Transverse momentum distribution of muons in the final state taking
  into account a minimum cut of $\pt(\mu^{\pm})>$~3 (left) 
  and $\pt(\mu^{\pm})>$~15~GeV (right) and also additional kinematic cuts 
  of $|\eta(\mu^{\pm})|<$~2.5, and $M_{X}=$~[1.07,1000.00]~GeV 
  in both cases. The distributions are obtained with \lpair using SY (solid line) and 
  SU (single-dashed line) structure functions,
  and with \pptoll where the $k_T$-factorization approach was implemented (double-dashed line).
  }
\end{figure}
\begin{figure}
  \includegraphics[width=.5\textwidth]{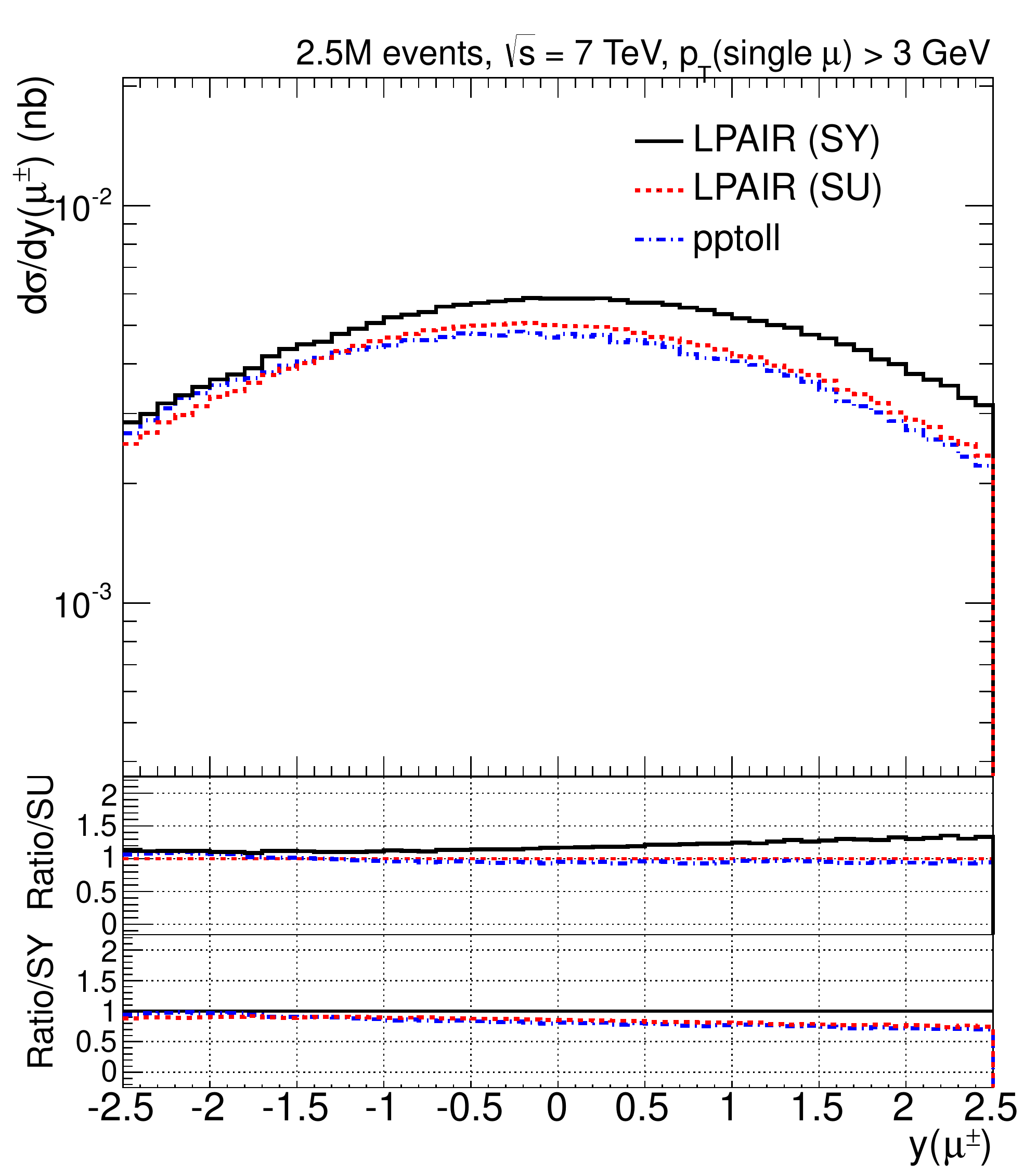}
  \includegraphics[width=.5\textwidth]{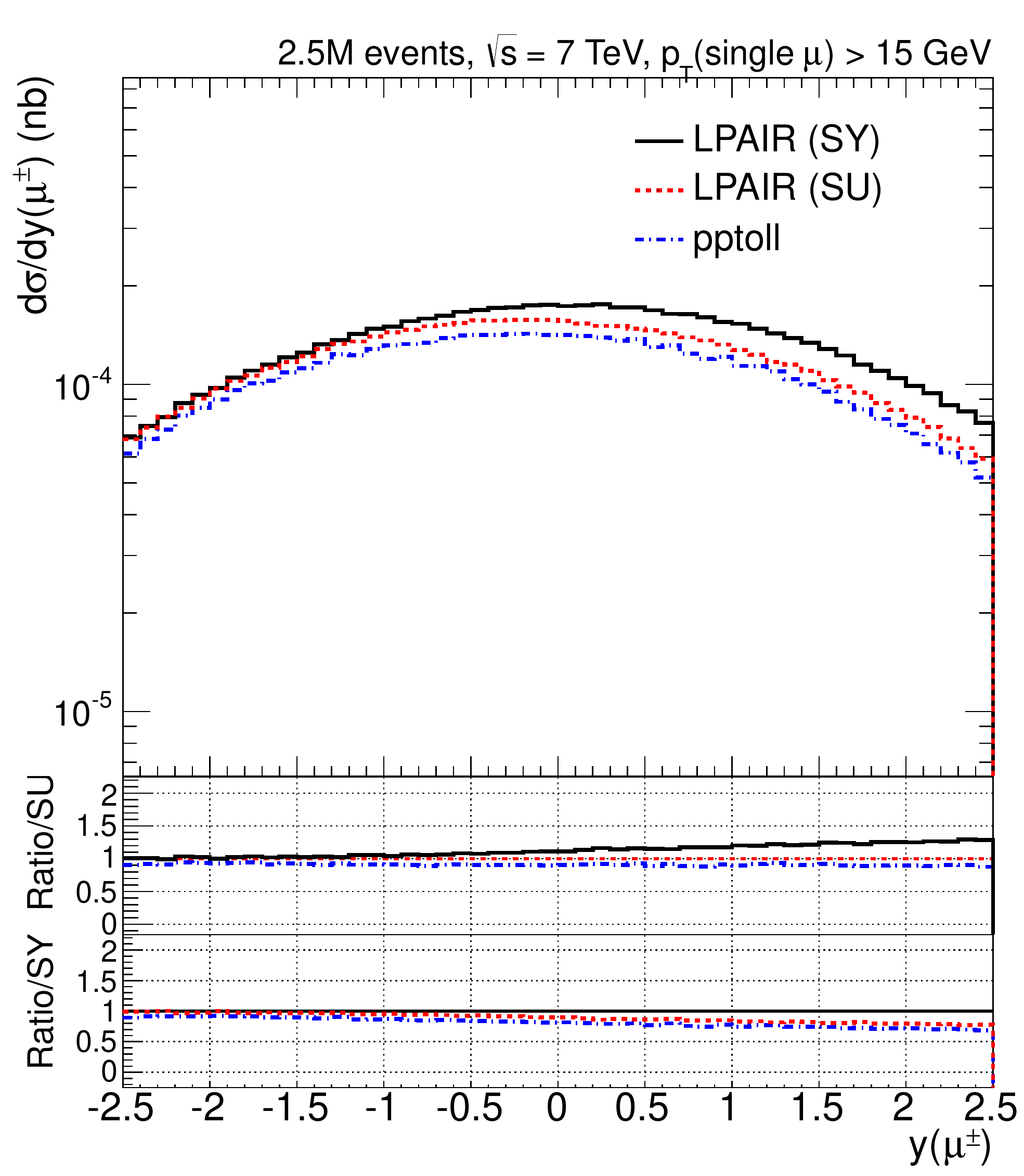}
  \caption{\label{etas}
  Rapidity distribution of dimuons in the final state taking into account
   $\pt(\mu^{\pm})>$~3~GeV (left) and $\pt(\mu^{\pm})>$~15~GeV (right) 
   and also additional kinematic cuts 
  of $|\eta(\mu^{\pm})|<$~2.5, and $M_{X}=$~[1.07,1000.00]~GeV 
  in both cases. The distributions are obtained with \lpair using SY (solid line) and 
  SU (single-dashed line) structure functions,
  and with \pptoll where the $k_T$-factorization approach was implemented (double-dashed line).
  }
\end{figure}
\begin{figure}
  \includegraphics[width=.5\textwidth]{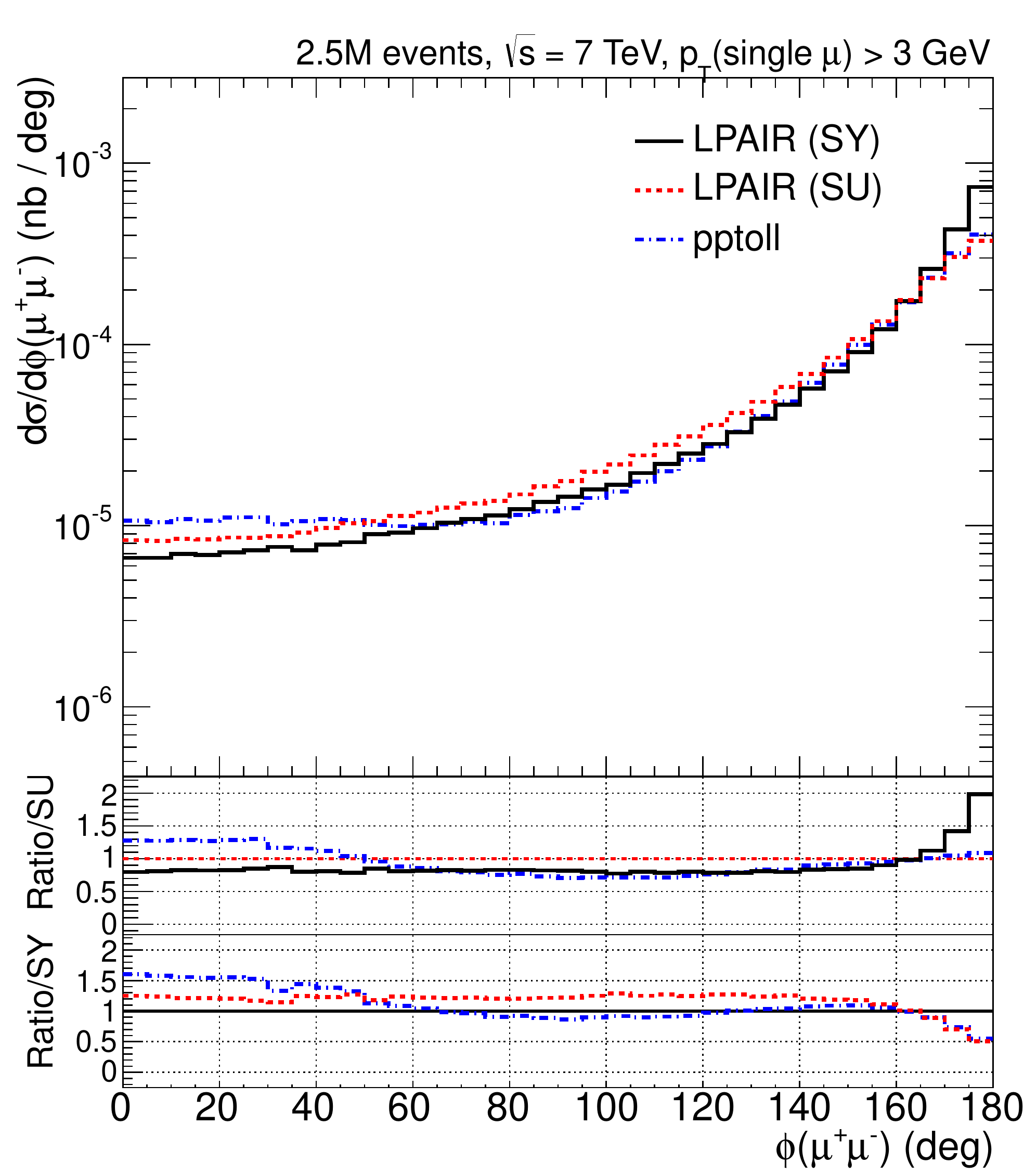}
  \includegraphics[width=.5\textwidth]{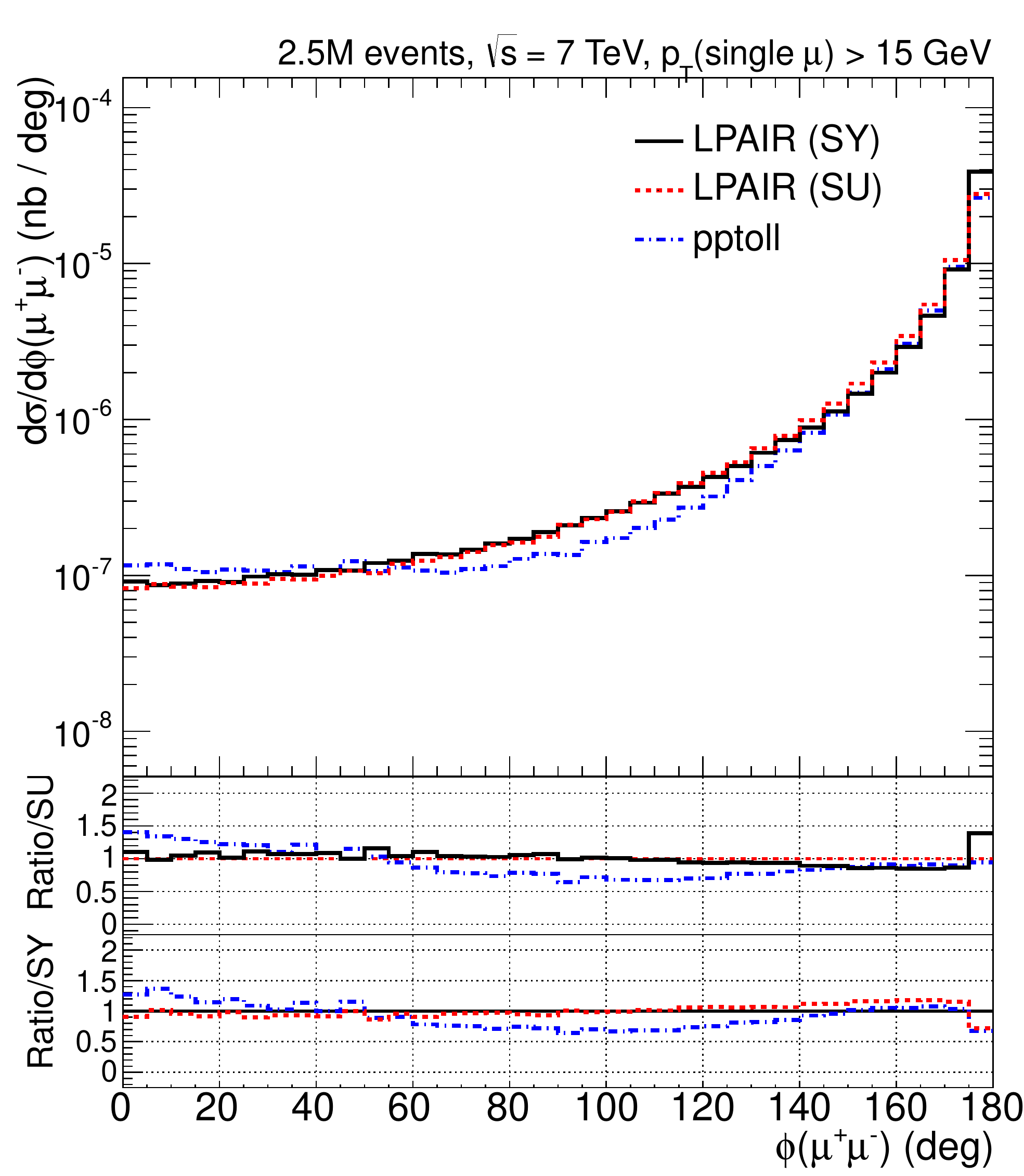}
  \caption{\label{phis}
  $\Delta\phi(\mu^{+}\mu^{-})$ distribution of muon pairs in the final state 
  taking into account $\pt(\mu^{\pm})>$~3~GeV (left) and 
  $\pt(\mu^{\pm})>$~15~GeV (right) and also additional kinematic cuts 
  of $|\eta(\mu^{\pm})|<$~2.5, and 
  $M_{X}=$~[1.07,1000.00]~GeV in both cases. The distributions are 
  obtained with \lpair using SY (solid line) and SU (single-dashed line) 
  structure functions,
  and with \pptoll where the $k_T$-factorization approach was implemented (double-dashed line).
  }
\end{figure}

\begin{figure}
  \includegraphics[width=.5\textwidth]{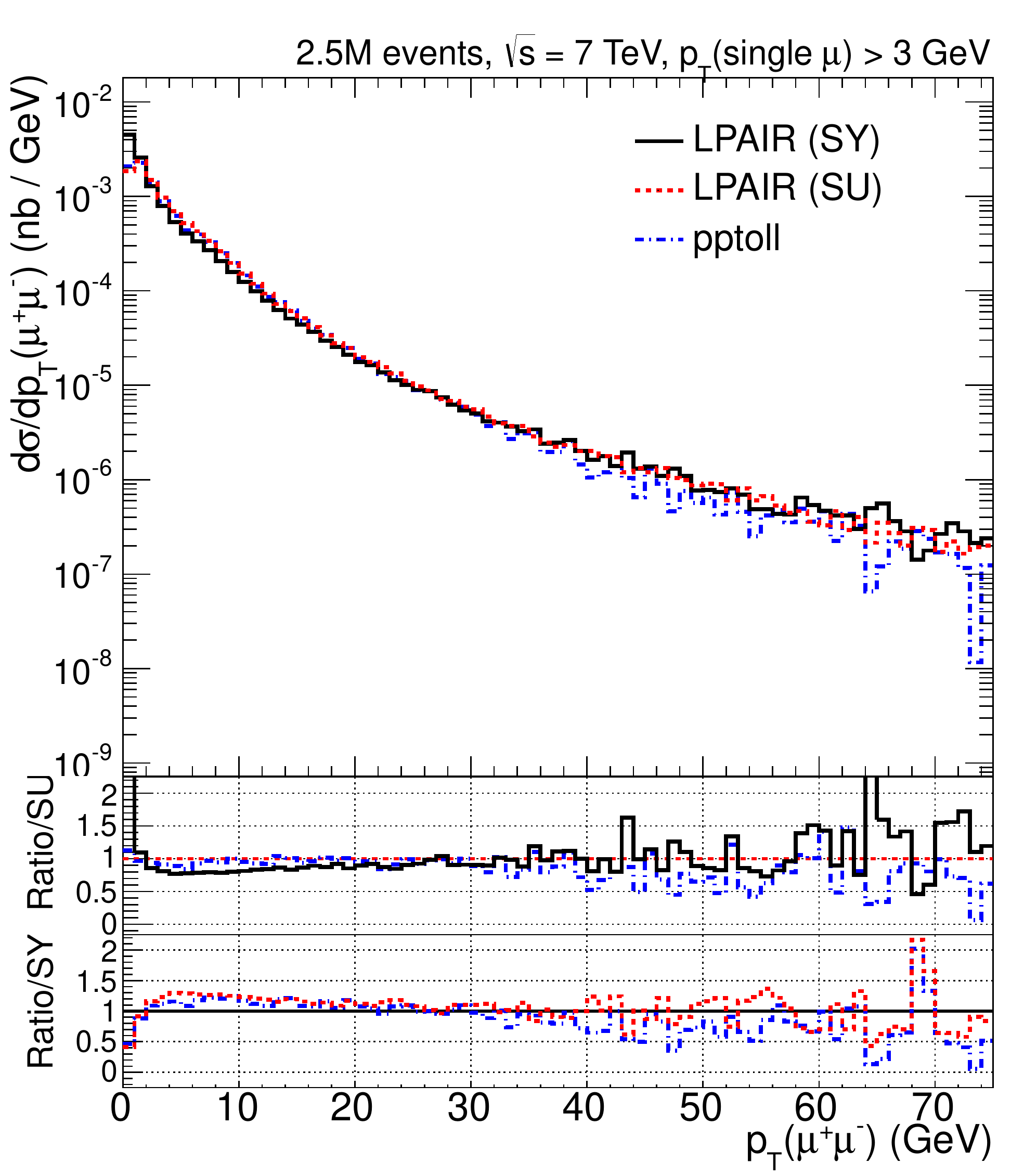}
  \includegraphics[width=.5\textwidth]{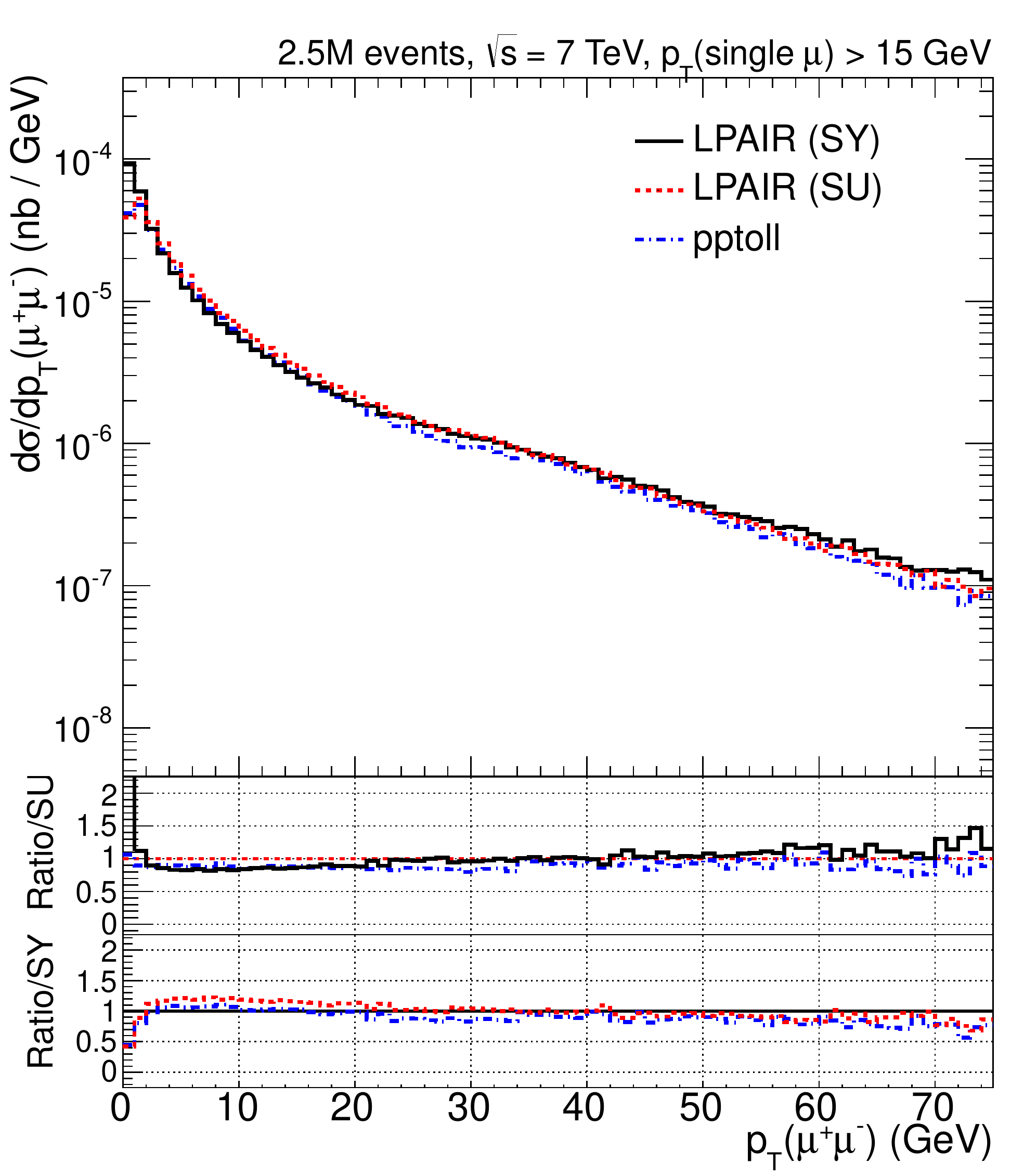}
  \caption{\label{pairs}
  Transverse momentum distribution of dimuons taking into account
   $\pt(\mu^{\pm})>$~3~GeV (left) and $\pt(\mu^{\pm})>$~15~GeV (right) 
   and also additional kinematic cuts of 
   $|\eta(\mu^{\pm})|<$~2.5, and $M_{X}=$~[1.07,1000.00]~GeV 
  in both cases. The distributions are obtained with \lpair using SY 
  (solid line) and SU (single-dashed line) structure functions,
  and with \pptoll where the $k_T$-factorization approach was implemented (double-dashed line).
  }
\end{figure}
\begin{figure}
  \includegraphics[width=.5\textwidth]{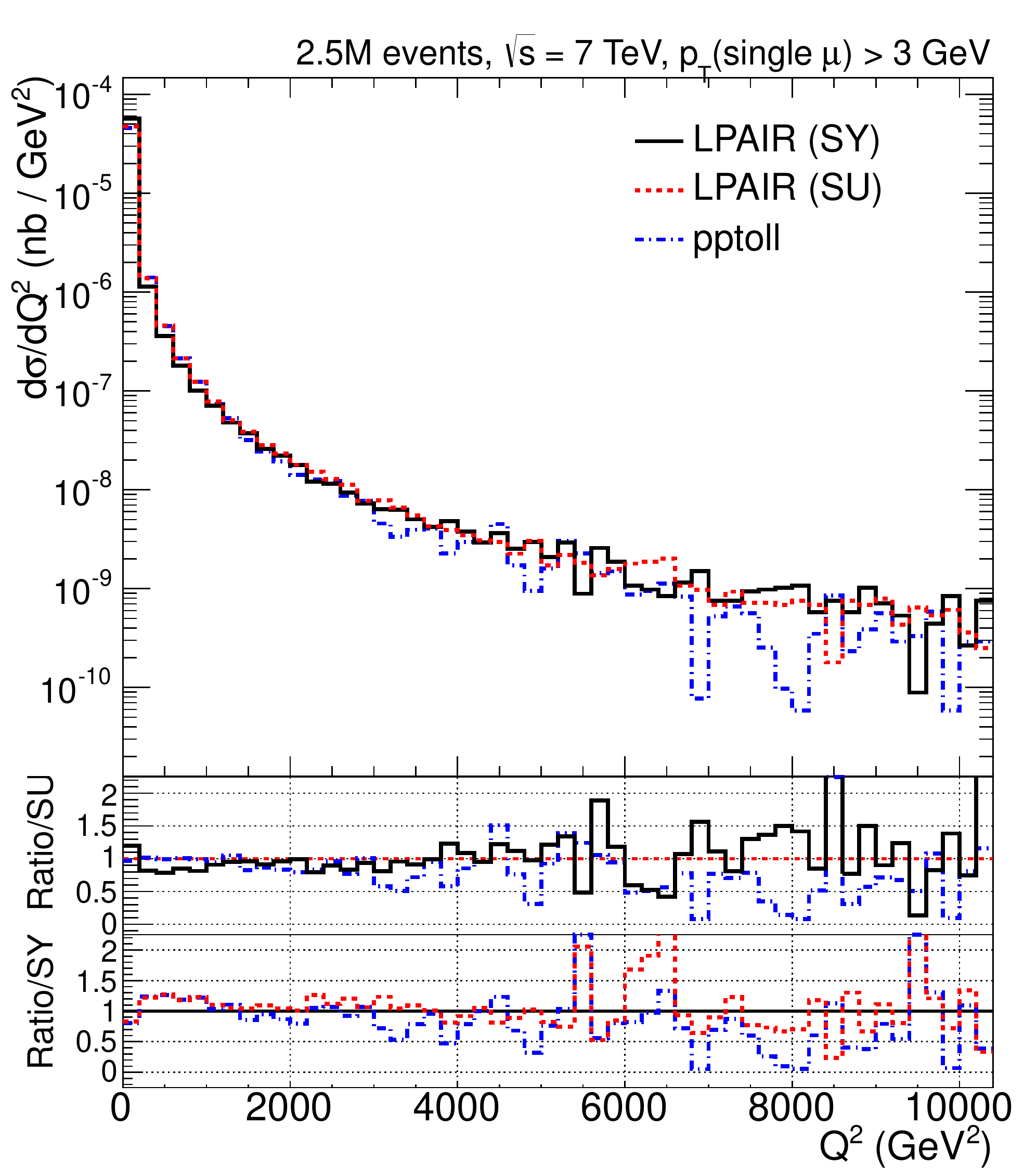}
  \includegraphics[width=.5\textwidth]{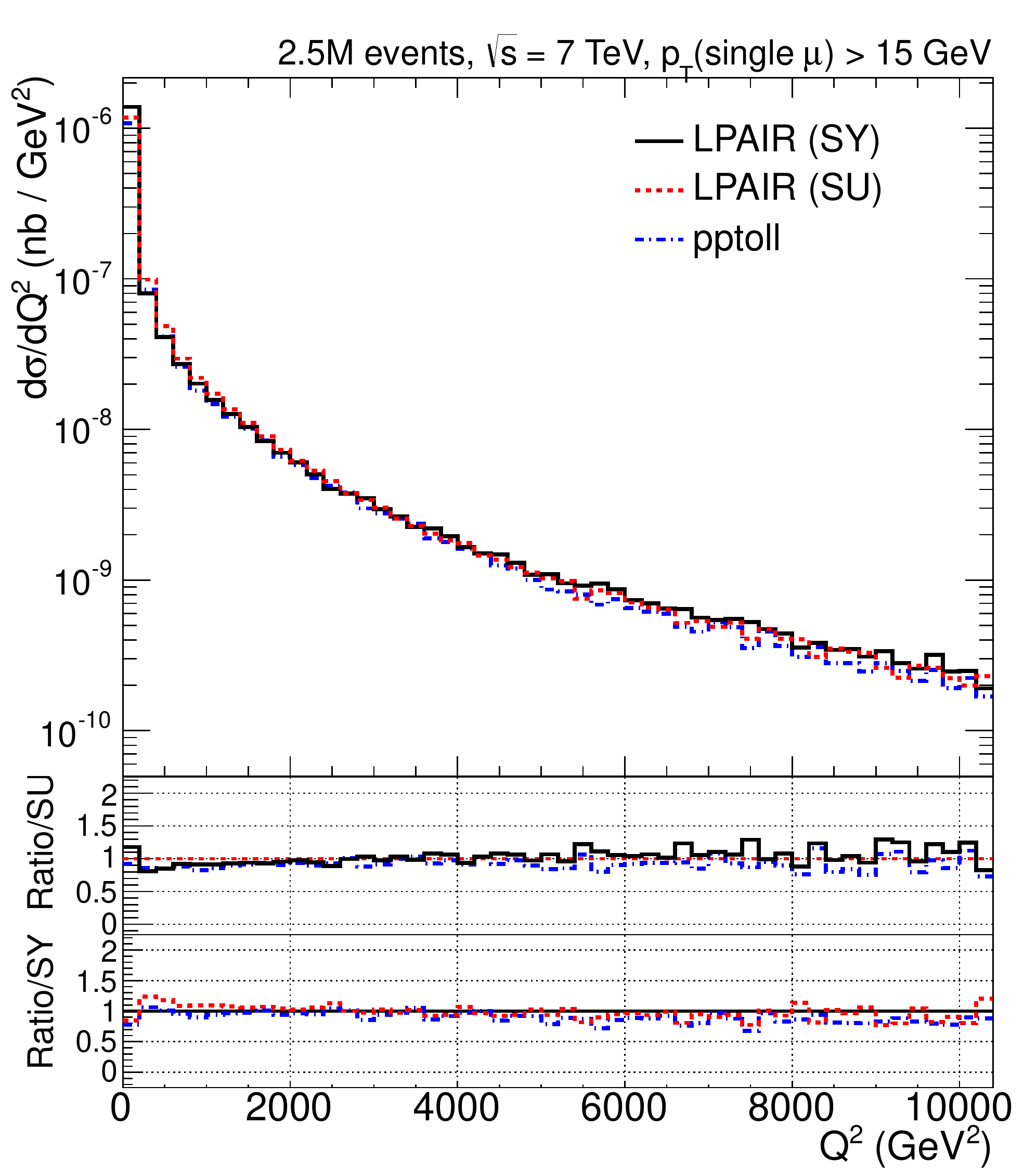}
  \caption{\label{q2plots}
  Photon virtuality distribution taking into account the cuts $\pt(\mu^{\pm})>$~3~GeV 
  (left) and $\pt(\mu^{\pm})>$~15~GeV (right) and also additional 
  kinematic cuts 
  $|\eta(\mu^{\pm})|<$~2.5, and $M_{X}=$~[1.07,1000.00]~GeV 
  in both cases. The distributions are obtained with \lpair using SY 
  (solid line) and SU (single-dashed line) structure functions,
  and with \pptoll where the $k_T$-factorization approach was implemented (double-dashed line).
  }
\end{figure}

We are interested in better understanding the $p_{T\textrm{sum}}$ spectra,
which can be nicely studied within the $k_T$-factorization
approach proposed in this paper, but cannot be addressed in simple EPA when transverse momenta 
of photons are neglected. Therefore, we wish to understand the correlation 
of the $p_{T\textrm{sum}}$ with other kinematical variables. 
In Fig.~\ref{2dplots} we show some examples of such two-dimensional 
correlations.
In the upper plots, the distribution in $M_{X}$ and $p_{T\textrm{sum}}$ is shown
for $p_T >$ 3 GeV (left) and for $p_T >$ 15 GeV (right).
We observe that the shape in $M_X$ strongly depends on $p_{T\textrm{sum}}$.
At lower $p_{T\textrm{sum}}$ the $M_X$-dependence is very sharp, while at
higher $p_{T\textrm{sum}}$ the dependence is much weaker.

The lower figures show distribution in $(p_{T\textrm{sum}},\Delta\phi)$.
One can observe clearly back-to-back type of correlations at
small $p_{T\textrm{sum}}$ and almost complete decorellation for large $p_{T\textrm{sum}}$
in the case of $p_T >$ 3 GeV (left).
The case of $p_T >$ 15 GeV (right) is, however, much more complicated.
One observes some areas which are forbidden due to kinematics.

\begin{figure}
  \includegraphics[width=0.49\textwidth]{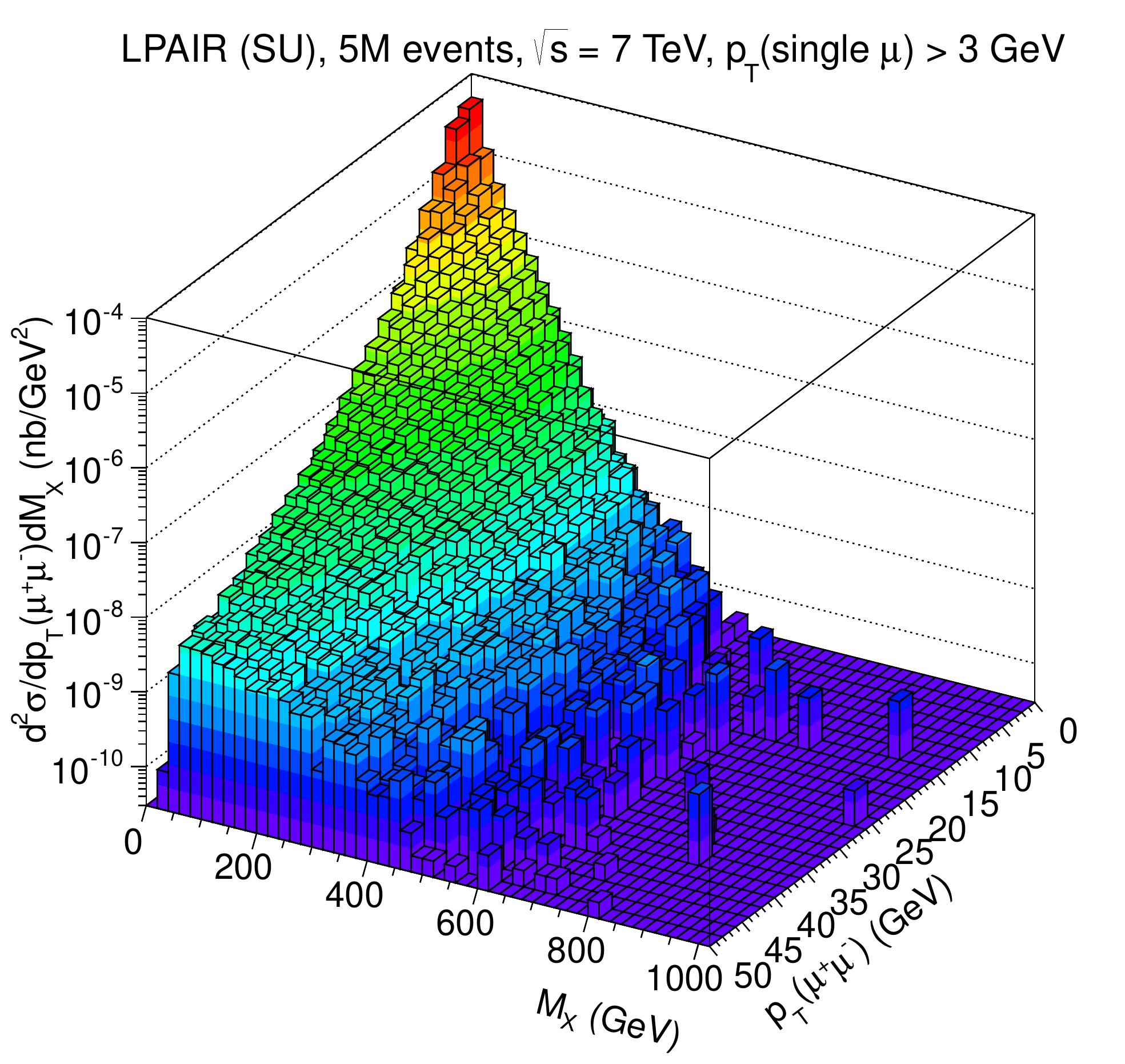}
  \includegraphics[width=0.49\textwidth]{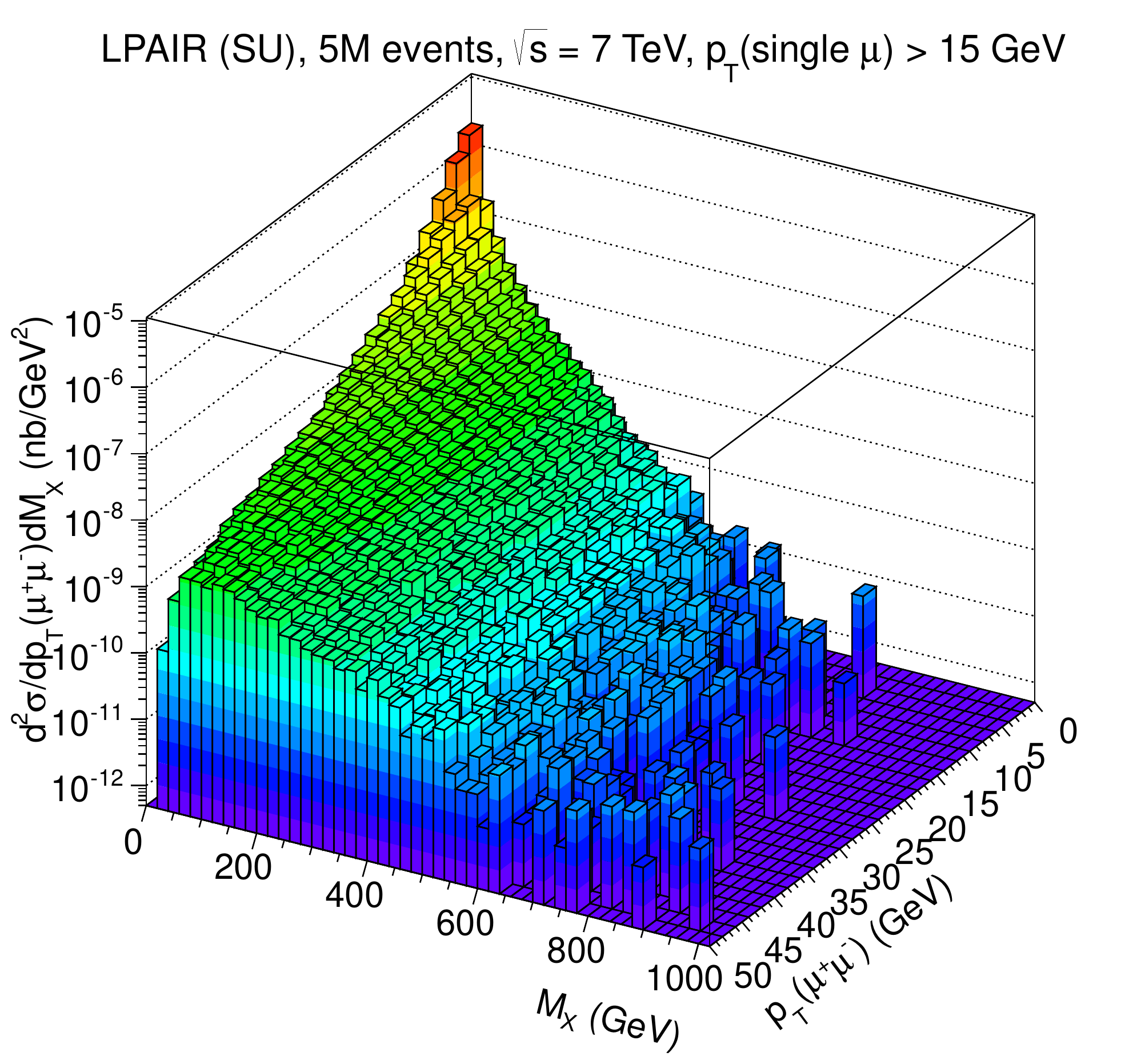}
  \includegraphics[width=0.49\textwidth]{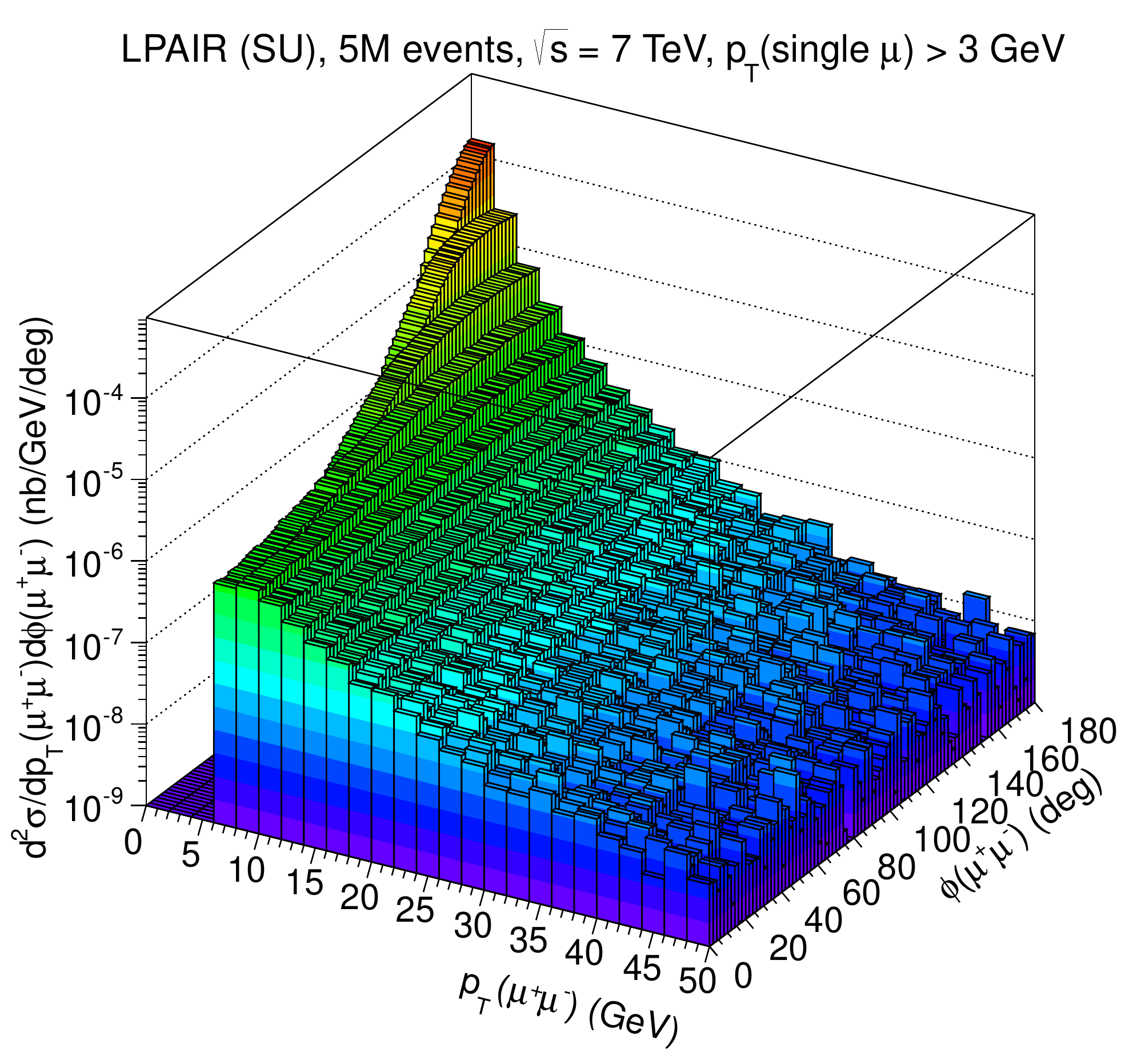}
  \includegraphics[width=0.49\textwidth]{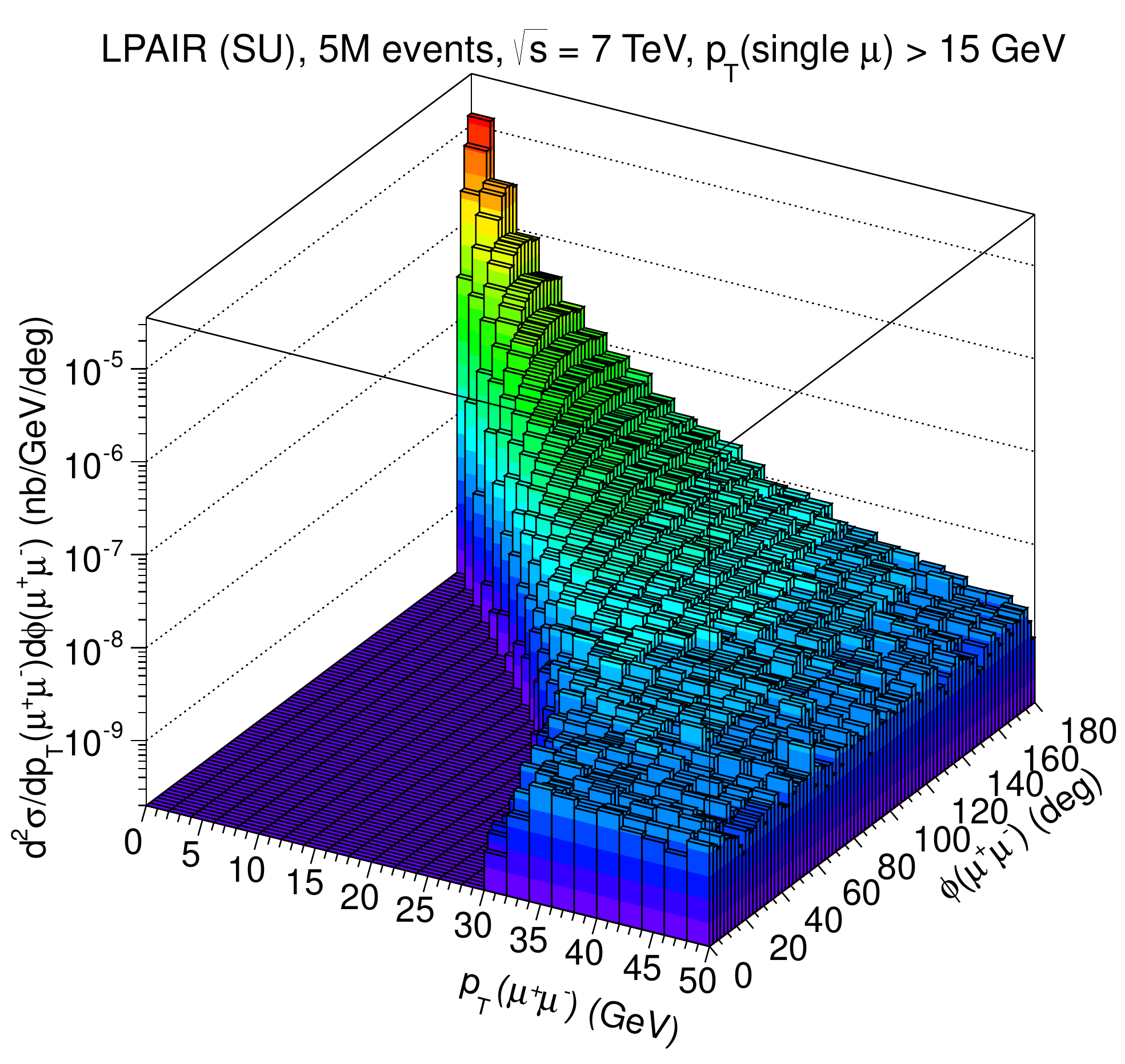}
  \caption{\label{2dplots}
  Different two-dimensional correlation of $p_{T\textrm{sum}}$ with $M_X$ (upper
  plots) and with $\Delta \phi$ (lower plots) for $p_T >$ 3 GeV (left side)
  and for $p_T >$ 15 GeV (right side) obtained with SU parametrization.
  In addition, we apply in both cases kinematical cuts of $|\eta(\mu^{\pm})|<$~2.5,
  and $M_{X}=$~[1.07,1000.00]~GeV.
  }
\end{figure}

\section{LPAIR vs. $k_{T}$-factorization approach}
\label{sec:compare}

For a complete overview of the physics related to the two-photon 
production of dimuons, we compare results obtained with the new
$k_t$-factorization approach (as described in
Sec.~\ref{sec:kt-approach}) with the \lpair code.
In case of elastic-elastic events we observe good agreement within the kinematic cuts used in this study
although the agreement gets somewhat worse when the $\pt$-cutoff gets larger.

In the following we concentrate on distributions for elastic-inelastic events, in which 
one proton dissociates while the other one stays intact.
We now turn to a brief discussion of the differences between 
the $k_T$-factorization and \lpair results. In the following, 
for the purpose of comparison, we shall in both approaches 
use only the $F_2$ structure function. 
First of all, the $F_1$ structure is not so well known at low scales 
and, secondly, it is only $F_2$ which appears in the high-energy limit.

We begin with the comparison of the $p_{T\textrm{sum}}$ distribution, shown in  
Figure~\ref{pairs}, for both the low-$p_T$ and high-$p_T$ cuts.
We observe some deviations in the shape of the distribution, especially
for the high-$p_T$ cut, where there appears to be an overprediction of
large-$p_{T\textrm{sum}}$ pairs, and a deficit at low $p_{T\textrm{sum}}$ in 
the $k_T$-factorization approach with respect to the \lpair results. 
This difference must be ascribed to the use of high-energy kinematics 
in the $k_T$-factorization approach.

Firstly, recall that due to experimental demands we integrate up 
to a very large mass of the excited system $M_X = 1 \, \mathrm{TeV}$. 
It is obvious that at such large $M_X$, the ''small'' Sudakov component 
of the ''inelastic photon'' [Eq.~(\ref{Sudakov_parameters})] is 
no longer negligible with respect to the momentum fraction $x$.
Among other things, this entails that momentum fractions $x_{1,2}$ cannot be
calculated from the muon kinematics alone, but must be calculated taking
the full final state into account. 

Secondly, in the most general case, the fluxes of longitudinal and 
transverse photons will differ \cite{budnev}, and a more involved
decomposition in terms of the full density matrix of photons must be adopted.

Notice that these deviations are concentrated in a kinematical region of large 
longitudinal momentum transfers. In impact parameter space, this would
correspond to a domain of intermediate to small impact parameters, where
absorptive effects are expected to be important, and the theoretical 
uncertainty associated with them is arguably higher than 
the here observed deviations between the two approaches. 
 
In the plots shown in Figs.~\ref{pts}--\ref{etas}
and \ref{masses}, we present other kinematic distributions
for both the $k_T$-factorization and \lpair approaches. 
The similar comments as above apply as far as deviations between the two
approaches are concerned.

\begin{figure}
  \centering
  \includegraphics[width=.49\textwidth]{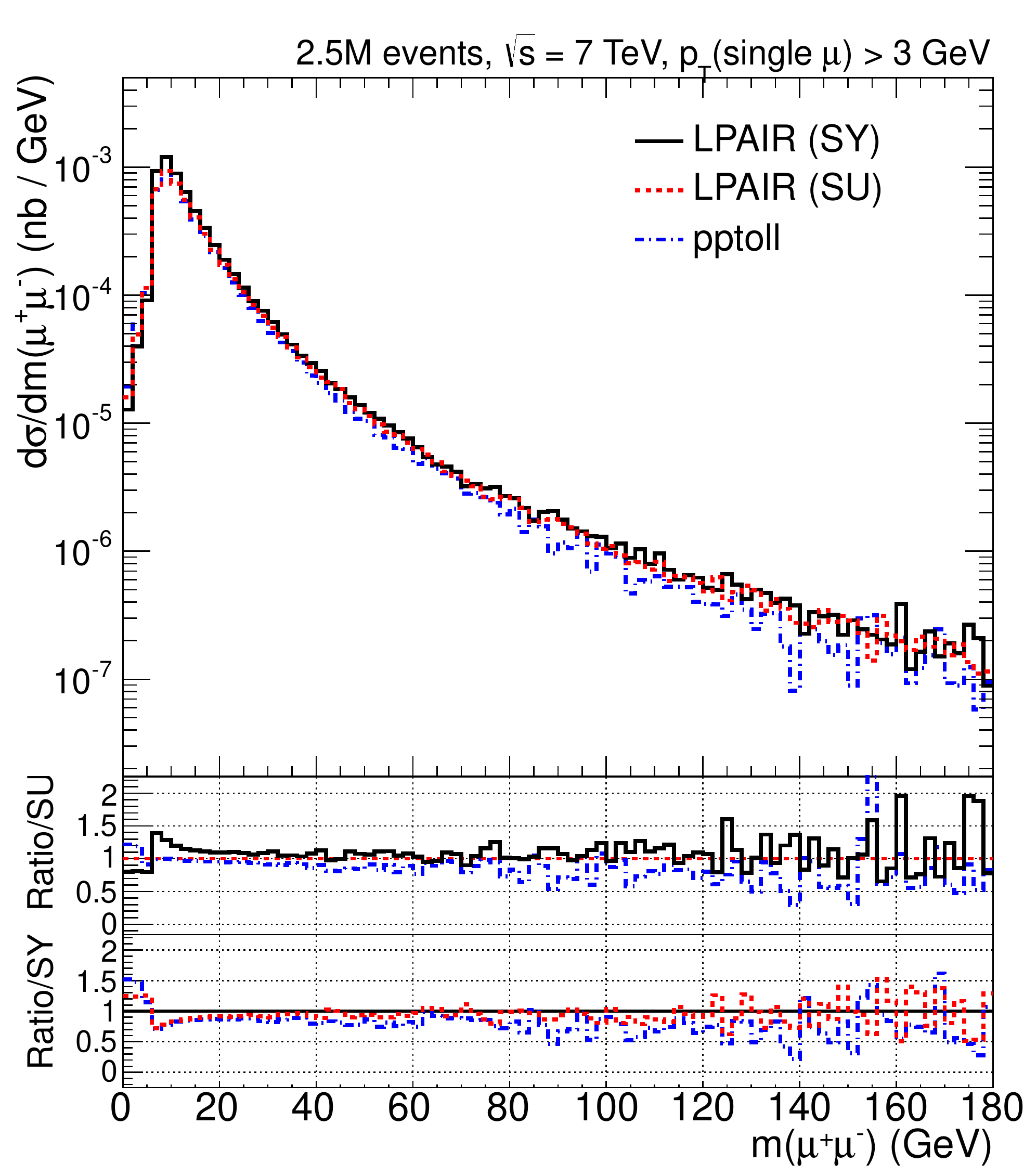}
  \includegraphics[width=.49\textwidth]{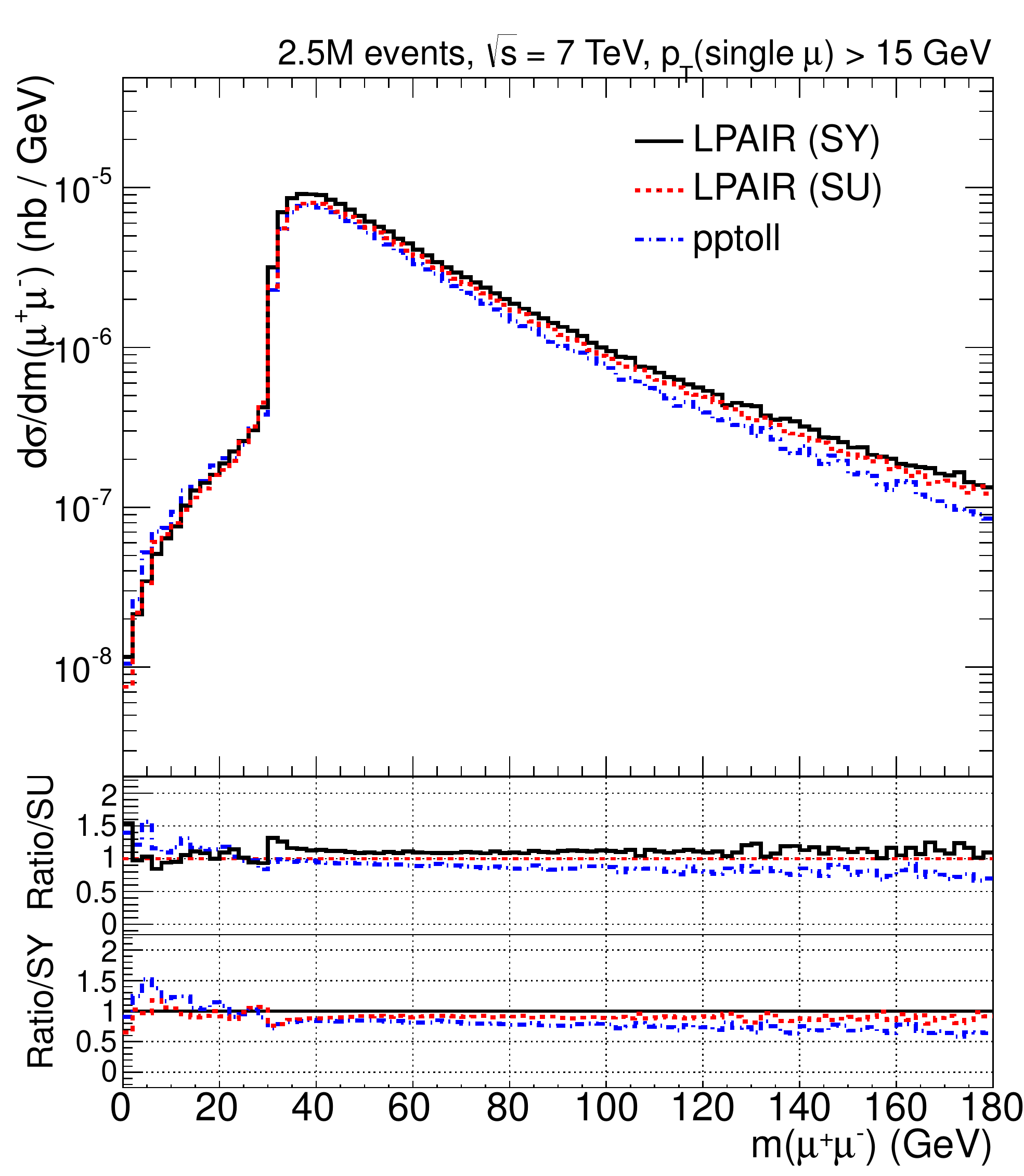}
  \caption{\label{masses}	
  Invariant mass distributions for muon pairs obtained with \lpair (solid and single-dashed line) in 
  comparison to the $k_T$-factorization predictions (double-dashed line) 
  for the interaction of off-shell photons as described 
  in Sec.~\ref{sec:kt-approach} taking into account kinematics cuts
  of $p_T>$~3~GeV (left column) and $p_T>$~15~GeV (right column) for 
  inelastic collisions.
  The histograms are the results obtained with \lpair using SY (solid line) and 
  SU (single-dashed line) structure functions,
  and with \pptoll implementing the $k_T$ factorization algorithm (double-dashed line).
  All distributions are obtained with the kinematic cuts 
  of $|\eta(\mu^{\pm})|<$~2.5, and $M_{X}=$~[1.07,1000.00]~GeV.
  }
\end{figure}

\section{Conclusions}
\label{concl}

In the present paper we have reviewed the production of muon pairs via the 
photon-photon fusion in proton-proton collisions at the LHC. All types of 
processes (elastic-elastic, elastic-inelastic, inelastic-elastic and 
inelastic-inelastic) have been discussed. We have performed the calculation with 
the code \lpair supplemented by the proton structure functions. The related 
uncertainties due to lack of precise knowledge of the latter objects have been
studied and discussed.

We have also proposed a new, somewhat simplified, method to calculate the 
processes in terms of photon unintegrated distributions that depend on photon 
longitudinal fraction and its transverse momentum. The results of the 
calculation have been compared with the results obtained with the \lpair code.
Rather good agreement between the two approaches has 
been achieved. The formalism proposed by us can be used not only for 
calculating integrated cross section or single muon distribution, but 
is especially well suited for efficient calculation of correlation observables.
As a results, the $k_{T}$-factorization approach can be extend to allow the 
study of the absorption corrections in inelastic collisions.

We have calculated distributions in dimuon invariant mass, in transverse 
momentum of the muon pair and in relative azimuthal angle between outgoing 
muons. The calculation has been performed separately for different components
and different proton structure functions from the literature. 
If the cuts on muon transverse momenta are imposed all 
components (elastic-elastic, 
elastic-inelastic, inelastic-elastic, inelastic-inelastic) give similar 
contribution. While the elastic-elastic contribution is well under control,
the inelastic contributions are subjected to uncertainties of the order of 
20\% or even more for some regions of correlation observables.
We have now at our disposal a good tool to calculate dileptons as 
a background to many other Standard Model processes. In many cases such 
a background was not even estimated. Similar method could be used 
for other photon-initiated processes.

We have shown that different regions of phase space are sensitive to
structure functions in different ranges of $x$ (longitudinal momentum 
fraction) and $Q^2$ (scale).
At present, there is no one set of structure functions in the literature
which correctly treats all regions of $x$ and $Q^2$.
We have used different sets of proton structure functions from the literature
and quantified uncertainties for cross section predictions
for $\mu^+ \mu^-$ production. They are typically of the order of 20\%--30\%,
and, although a significant sensitivity to the parton distribution 
function is observed, it cannot explain high-$\pt$ results obtained 
by the CMS Collaboration.

In case of the semi-exclusive events at large $\pt$ of the pair absorption effects must be included in the 
calculation. We do not need to mention that the calculation of such effects
in terms of kinematic variables is not an easy task. While some early discussions of 
absorptive effects in lepton pair production do exist
\cite{Khoze:2000db,HarlandLang:2012qz}, these studies concentrate on an entirely 
different kinematic region than in this work. It therefore 
needs to be understood whatever such absorption effects could 
be \emph{extracted} by comparison of the calculated cross sections 
with the measured ones. This subject will be studied in more
detail in the future.

\acknowledgments

We are indebted to Rafa{\l} Maciu{\l}a for help in preparing a Monte Carlo 
version of the code calculating lepton production in the $k_T$-factorization 
approach. The authors strongly acknowledge the support by the dedicated bilateral project 
between the FNRS in Belgium and PAN in Poland.
GGS thanks the partial support by CNPq and Capes, Brazil.
This work was partially supported by the Polish MNiSW grant 
DEC-2011/01/B/ST2/04535 as well as by the Centre for Innovation and Transfer 
of Natural Sciences and Engineering Knowledge in Rzesz\'ow.



\begin{thebibliography}{99}

\bibitem{Budnev:1973tz}
  V.~M.~Budnev, I.~F.~Ginzburg, G.~V.~Meledin and V.~G.~Serbo,
  \emph{The process $pp\to pp e^{+}e^{-}$ and the possibility of its calculation by means of quantum electrodynamics only},
  \emph{Nucl. Phys.} {\bf B63} (1973) 519.

\bibitem{budnev} V.M. Budnev, I.F. Ginzburg, G.V. Meledin, V.G. Serbo,
  \emph{The Two photon particle production mechanism. Physical problems. Applications. Equivalent photon approximation}, 
  \emph{Phys. Rept.} {\bf 15} (1975) 181.

\bibitem{SU2000}
  A. Szczurek and V. Uleshchenko, 
  \emph{Nonpartonic components in the nucleon structure functions at small $Q^{2}$ in the broad range of $x$},
  \emph{Eur. Phys.} {\bf C12} (2000) 663,
  arxiv:hep-ph/9904288.

\bibitem{SU2000_GSR}
  A. Szczurek and V. Uleshchenko,
  \emph{On the range of validity of the QCD improved parton model},
  \emph{Phys. Lett.} {\bf B475} (2000) 120,
  arxiv:hep-ph/9911467.

\bibitem{Fiore:2002re} 
  R.~Fiore, A.~Flachi, L.~L.~Jenkovszky, A.~I.~Lengyel and V.~K.~Magas,
  \emph{Explicit model realizing parton hadron duality},
  \emph{Eur. Phys. J.} {\bf A15}, 505 (2002),
  arxiv:hep-ph/0206027.

\bibitem{Maciula:2013wg}
  R.~Maciu\l a and A.~Szczurek,
  \emph{Open charm production at the LHC - $k_{t}$-factorization approach}
  \emph{Phys. Rev.} {\bf D87} (2013) 094022,
  arxiv:1301.3033 [hep-ph].

\bibitem{lpair}
  S.P. Baranov, O., D\"unger, H., Shooshtari, J.A.M. Vermaseren;
  \emph{\lpair: A generator for lepton pair production},
  \emph{Proceedings of Physics at HERA} {\bf 3} 1478.

\bibitem{vermaseren}
  J.A.M. Vermaseren, 
  \emph{Two Photon Processes at Very High-Energies}, 
  \emph{Nucl. Phys.} {\bf B229} (1983) 347.

\bibitem{Bartos:2001jz} 
  E.~Bartos, S.~R.~Gevorkyan, E.~A.~Kuraev and N.~N.~Nikolaev,
  \emph{The lepton pair production in heavy ion collisions in perturbation theory},
  \emph{Phys. Rev.} {\bf A66}, 042720 (2002),
  arxiv:hep-ph/0109281.

\bibitem{Frolov:1970ij} 
  G.~V.~Frolov, V.~N.~Gribov and L.~N.~Lipatov,
  Phys.\ Lett.\ B {\bf 31}, 34 (1970) ; \\
  L.~N.~Lipatov and G.~V.~Frolov,
  Pisma Zh.\ Eksp.\ Teor.\ Fiz.\  {\bf 10}, 399 (1969).

\bibitem{Cheng:1970sz} 
  H.~Cheng and T.~T.~Wu,
  Phys.\ Rev.\ D {\bf 1}, 456 (1970).

\bibitem{Szczurek:2000pj} 
  A.~Szczurek, N.~N.~Nikolaev, W.~Sch\"afer and J.~Speth,
  Phys.\ Lett.\ B {\bf 500}, 254 (2001)
  [hep-ph/0011281].

\bibitem{vegas}
  G.P. Lepage,
  \emph{A New Algorithm for Adaptive Multidimensional Integration},
  \emph{J. Comp. Phys.} {\bf 27} (1978) 192.

\bibitem{LS2006}
  M. {\L}uszczak and A. Szczurek,
  \emph{Gluon transverse momenta and charm quark-antiquark pair production in $p\bar p$ collisions at Tevatron},
  \emph{Phys. Rev.} {\bf D73} (2006) 054028,
  arxiv:hep-ph/0512120.

\bibitem{MSS2011}
  R. Maciu{\l}a, A. Szczurek and G. \'Slipek, 
  \emph{Kinematical correlations of dielectrons from semileptonic decays of heavy mesons and Drell-Yan processes at BNL RHIC},
  \emph{Phys. Rev.} {\bf D83} (2011) 054014,
  arxiv:1011.4207 [hep-ph].

\bibitem{suri}
  A. Suri, D.R. Yennie, 
  \emph{The space-time phenomenology of photon absorption and inelastic electron scattering}, 
  \emph{Annals Phys.} {\bf 72} (1972) 243.

\bibitem{cms-papers} 
  CMS Collaboration,
  \emph{Exclusive photon-photon production of muon pairs in proton-proton collisions at $\sqrt{s}$~=~7~TeV},
  \emph{JHEP} {\bf 01} (2012) 052,
  arxiv:1111.5536 [hep-ex] ; \\
  CMS Collaboration,
  \emph{Search for exclusive or semi-exclusive photon pair production and observation of exclusive and semi-exclusive electron pair production in $pp$ collisions at $\sqrt{s}$~=~7~TeV},
  \emph{JHEP} {\bf 11} (2012) 080,
  arxiv:1209.1666 [hep-ex].

\bibitem{pythia}
  T. Sj\"ostrand, S. Mrenna, P. Z. Skands,
  \emph{PYTHIA 6.4 Physics and Manual}, 
  \emph{JHEP} {\bf 05} (2006) 026.

\bibitem{fsq-12-010}
  CMS Collaboration,
  \emph{Study of exclusive two-photon production of $W^{+}W^{-}$ in $pp$ collisions at $\sqrt{s}$~=~7~TeV and constraints on anomalous quartic gauge couplings},
  \emph{JHEP} {\bf 07} (2013) 116.

\bibitem{Khoze:2000db}
  V.A. Khoze, A.D. Martin, R. Orava, M.G. Ryskin,
  \emph{Luminosity monitors at the LHC},
  \emph{Eur. Phys. J.} {\bf C19} (2001) 313,
  arxiv:hep-ph/0010163.

\bibitem{HarlandLang:2012qz}
  L.A. Harland-Lang, V.A. Khoze, M.G. Ryskin, W.J. Stirling,
  \emph{The phenomenology of Central Exclusive Production at Hadron Colliders},
  \emph{Eur. Phys. J.} {\bf C72} (2012) 2110,
  arXiv:1204.4803[hep-ph]

\end{thebibliography}
\end{document}